\documentclass[notitlepage,
 twocolumn,
superscriptaddress,
nofootinbib,
 amsmath,amssymb,
 aps,
 prb]{revtex4-2}
\usepackage{amssymb,graphicx,mathtools,dsfont,bbold, natbib}
\usepackage[dvipsnames]{xcolor}
\usepackage[T1]{fontenc}
\usepackage{soul}
\usepackage{bibunits}
\usepackage{dcolumn}% Align table columns on decimal point
\usepackage{bm}% bold math
\usepackage{color}
\usepackage{physics}
\usepackage{dsfont}
\usepackage{adjustbox}
\usepackage{soul}
\usepackage{multirow}
\usepackage{rotating} % Rotating table
\usepackage{comment}
\usepackage{hyperref}
\hypersetup{citecolor=red,colorlinks=true,urlcolor=blue}
\usepackage{amsmath}
\usepackage{graphicx}

\usepackage{booktabs}

\newcommand{\bfx}{\mathbf{x}}
\newcommand{\bfr}{\mathbf{r}}

\newcommand{\bfl}{\boldsymbol{\lambda}}
\newcommand{\bfm}{\boldsymbol{\mu}}

\usepackage[utf8]{inputenc}

\begin{document}

%\preprint{}

\title{%Quantum hard-rods\\
%Form and structure factors of quantum hard rods in one dimension: an analytical and semi-analytical study\\
%Analytical insights into form and structure factors of quantum hard rods \\
Dynamic structure factor of quantum hard rods from exact form-factors}


\author{Stanisław Kiedrzyński}
\affiliation{Faculty of Physics, University of Warsaw, Pasteura 5, 02-093 Warsaw, Poland}
%\thanks{}
\author{Emilia Witkowska}
\affiliation{Institute of Physics PAS, Aleja Lotników 32/46, 02-668 Warsaw, Poland}
\author{Miłosz Panfil}
\affiliation{Faculty of Physics, University of Warsaw, Pasteura 5, 02-093 Warsaw, Poland}


\date{\today}

\begin{abstract}
We study the quantum hard-rods model and obtain compact analytical expressions for density form factors, 
and a semi-analytical treatment for dynamic and static structure factors calculations, greatly reducing computational complexity. 
We identify conditions under which these form factors vanish and analyze real-space correlations, confirming the model's Tomonaga–Luttinger liquid behavior.
The results reveal universal features of low energy physics of a gapless quantum fluid and its relation to Luttinger liquid theory, providing precise benchmarks for numerical simulations. 
This work establishes quantum hard rods as an important testbed for theories of strongly correlated one-dimensional systems.
\end{abstract}

%\keywords{}

\maketitle

\section{Introduction}

Strongly correlated systems are a subject of intense research in condensed matter physics, offering insights into complex quantum phenomena~\cite{Quintanilla_2009}. 
In particular, one-dimensional many-body systems provide a unique laboratory for testing theories, often allowing for analytical solutions~\cite{Giamarchi2004, Franchini2016}. 
A canonical model for a one-dimensional gas in that respect is the Lieb-Liniger model~\cite{1963_Lieb_PR_130_2} which due to its quantum integrability and experimental realizations has led to a plethora of results in the field of equilibrium and non-equilibrium physics~\cite{Paredes2004,Kinoshita2004,PhysRevLett.115.085301,PhysRevLett.91.250402,PhysRevLett.121.160603,PhysRevLett.122.090601,quasi_1d}. %Bouchoule_2022 
The \emph{quantum hard rods model} is a similar but less explored system~\cite{Lenard64, Lenard66}. Its quantum integrability yields analytical solvability for an exact many-body wavefunction and excitation spectra, known thanks to the Bethe ansatz~\cite{1940705, 10.1063/1.1665585}. 
However, its dynamic properties have been weakly addressed with exact analytical methods, although numerical investigations are performed. 
This is in striking contrast to its classical counterpart, the gas of hard rods, explored extensively since the seminal paper of Tonks~\cite{PhysRev.50.955} through studies of correlation functions
~\cite{SalsburgZwanzigKirkwood1953,10.1063/1.1664471}, to modern developments in its nonequilibrium dynamics~\cite{Doyon2018_PRL_SolitonGases,DoyonSpohn2017_DomainWall,SinghDharSpohnKundu2024_JStatPhys_HardRods}. It provides also an exceptional example of an interacting system in which phenomena as complex as hydrodynamics can be derived from first principles~\cite{SpohnBOOK}.

The dynamic properties of the quantum hard rods model have been studied, mostly numerically, including the dynamic structure factor and off-diagonal ground-state properties \cite{PhysRevA.94.043627, PhysRevA.77.043632, PhysRevLett.100.020401}. 
These works provide the background for further, more detailed analyses, especially in the context of dynamic correlation functions. 
An experimental implementation has yet to be achieved; nonetheless, several proposals—such as those involving Rydberg atom systems—present a promising path forward~\cite{PhysRevLett.95.185302,yu2025exactsolutionluttingerliquid}.
Apart of these advances, the literature remains shortage of exact analytical results that would enable a more detailed understanding of the model and allow for more definitive conclusions. Such results are crucial for rigorously analyzing correlations, form factors, and dynamic properties, and for providing benchmarks to guide both theoretical developments and experimental studies.

In this work, we investigate a system of $N$ quantum particles interacting via the quantum hard-rod potential in one dimension. % of size $L$.
The particles undergo interactions and are not allowed to approach each other closer than a distance $a$, which is interpreted as the interaction range or the rod length. 
An exchange of particle momenta takes place due to elastic collisions, leading to non-trivial static and dynamic properties of the quantum hard rods gas.

Using the exact Bethe-ansatz solution~\cite{1940705}, we perform a detailed calculation of the form factors of the density operator between arbitrary eigenstates of the model. We derive a compact analytical expression for these form factors in terms of a Cauchy determinant, reducing their computational complexity from $(N!)^2$ to $N^3$.
This explicit analytic result enables us to analyze behavior of form factors, essential for the subsequent evaluation of the dynamic structure factor via the spectral sum, as well as the static structure factors.
We also analytically investigate the real-space correlation function. Our analysis confirms that the system falls within the Tomonaga–Luttinger (TL) liquid universality class. We further derive explicit analytical expressions for the asymptotic form of the correlation function, establishing a direct correspondence with predictions from Luttinger liquid theory. In the sparsely packed regime, the correlation function exhibits oscillatory behavior consistent with TL theory. In contrast, in the densely packed limit, $N$ equally spaced peaks emerge, reflecting the fact that fixing the position of one hard rod at the origin constrains the positions of the remaining rods to be centered at integer multiples of $a$.

The collection of analytical results obtained in this work constitutes a significant contribution to understanding the dynamic properties of strongly correlated one-dimensional systems. 
They provide accurate reference points for benchmarking numerical methods and form a solid foundation for further theoretical studies of one-dimensional quantum gases and their dynamics. Moreover, this work reinforces the role of the hard rods model as a benchmark system for testing general theories of quantum many-body physics.

\section{The Hamiltonian}

The Hamiltonian of the quantum hard rods is
\begin{equation}
	H = \sum_{j=1}^N \frac{\hat{p}_j^2}{2m} + \sum_{i<j} V_{\rm hr}(r_i - r_j),
    \label{eq:ham}
\end{equation}
where $\hat{p}_j$ and $r_j$ are the momentum operator and position of the $j$-th particle, and 
where the hard-rods potential reads
\begin{equation}
	V_{\rm hr}(r) = 
	\begin{cases}
		\infty, \qquad &|r| \leq a, \\
		0, \qquad &|r| > a,
	\end{cases}
\end{equation}
with rods length $a$.
In the following, we set $2m=1$ and $\hbar = 1$. 
We assume the system has length $L$. 
The quantum hard-rods model is solvable analytically by the coordinate Bethe ansatz and the exact many-body wavefunction takes the form of superposition of plane waves~\cite{Bethe1931}. 

We consider particles of either fermionic or bosonic statistics. The wavefunction for fermions is
\begin{equation} \label{wave_function_sum}
	\psi_F(\bfr| \bfl) = \frac{1}{\sqrt{\mathcal{N}}}\sum_{\sigma \in \mathcal{P}_N} (-1)^{|\sigma|} \psi_{\sigma}(\bfr|\bfl),
\end{equation}
where 
we introduced a compact notation for a set of positions $\bfr = \{r_1, \dots, r_N\}$ and rapidities $\bfl = \{\lambda_1, \dots, \lambda_N\}$, the latter are also called quasi-momenta in literature~\cite{1940705}. The normalization $\mathcal{N}$ depends on the boundary conditions, and we leave it unspecified for now.

In the many-body wave function (\ref{wave_function_sum}) the $\sigma$ summation runs over all permutations $\mathcal{P}_N$ of rapidities in the corresponding plane waves 
\begin{equation} \label{wave_function_perm}
	\psi_{\sigma}(\bfr| \bfl) = e^{i \sum_{j=1}^N \lambda_{\sigma_j} r_j - \frac{i}{2}\sum_{j>l}^N {\rm sgn}(r_j - r_l)\theta(\lambda_{\sigma_j} - \lambda_{\sigma_l})},
\end{equation}
with the phase shift $\theta(\lambda_j - \lambda_l) = (\lambda_j - \lambda_l) a$ found by solving the $2$-body problem~\cite{10.1063/1.1665585}. 
The simple form of the scattering phase shift leads to an expression for the wavefunction in the form of the Slater determinant
\begin{equation}
	\psi_F(\bfr| \bfl) = \frac{1}{\sqrt{\mathcal{N}}} \det\left(e^{i \lambda_j(r_l - (l-1)a)} \right),
    \label{wave_function_F}
\end{equation}
see Appendix~\ref{app:Slater} for the details of the computation.

For bosonic particles the wavefunction is
\begin{equation} \label{wave_function}
	\psi_B(\bfr | \bfl) = \prod_{j<l }{\rm sgn}(r_j - r_l) \times \psi_F(\bfr| \bfl),
\end{equation}
where the introduced factor ensures the correct symmetry of the wavefunction. The consequence of this relation is that the Pauli principle, the vanishing of the wave function when two coordinates coincide, holds also in the bosonic theory. 

The momentum and the energy of the eigenfunction $\psi_{F/B}$ take the standard form for galilean invariant theories,
\begin{equation}
	P_{\bfl} = \sum_{j=1}^N \lambda_j, \qquad E_{\bfl} = \sum_j^N \lambda_j^2.
\end{equation}
 
Imposing the periodic boundary condition leads to the quantization of the rapidities in the form of the Bethe equations
\begin{equation}
	\lambda_j = \frac{2\pi}{L} I_j + \frac{1}{L}\sum_{l \neq j}^N \theta(\lambda_j - \lambda_l),  \quad j = 1, \dots ,N,
\end{equation}
where $I_j$ are quantum numbers (integers for fermions and odd number of bosons, half-odd integers for even number of bosons). The simple form of the phase shift allows for an explicit solution
\begin{equation} \label{eq:Bethe}
	\lambda_j = \frac{2\pi}{L_{\rm f}} I_j - \frac{a P_{\bfl}}{L_{\rm f}},
\end{equation}
with the momentum expressed in terms of the quantum numbers and where $L_{\rm f} = L - Na$ is the free length accessible to $N$ particles of size $a$ on a line of length $L$.
\begin{equation}
 	P_{\bfl} =  \frac{2\pi}{L }\sum_{j=1}^N I_j.
   \label{eq:defP}
\end{equation}
From a direct inspection of the wavefunction it follows that the Pauli principle holds also for the quantum numbers. 
In the following we assume that every eigenstate of the theory can be described by an appropriate choice of the quantum numbers and the resulting eigenfunction~\eqref{wave_function_F} or \eqref{wave_function}.

The filling fraction, $\rho_0 a$ where $\rho_0 = N/L$ is the 1d density, provides a dimensionless parameter characterising the 'interaction' strength in the system. Another useful quantity is the Luttinger liquid parameter $K$ with the two simply related, $K = (1 - \rho_0 a)^2$~\cite{PhysRevLett.100.020401}. For the hard rods the Luttinger liquid is in the repulsive regime where $0\leq K\leq 1$, with $K=1$ for non-interacting case of hard-rods of zero length and $K=0$ for fully packed and strongly correlated system~\cite{GiamarchiBOOK}. This is reflected in the expression for the free volume, which in terms of the Luttinger parameter is $L_{\rm f} = \sqrt{K} L $.

The ground state of the theory, given the Pauli principle, is obtained by the most compact choice of quantum numbers around $0$. For bosonic theories and for odd numbers of fermions this leads to the unique ground state with the following quantum numbers
\begin{equation}
	I_{j}^{\rm gs} = - \frac{N+1}{2} + j,\qquad j = 1, \dots, N.
\end{equation}
It has zero momentum $P_{\rm gs} =0$ and energy $E_{\rm gs}/N = \pi^2(N^2 - 1)/(6L_{\rm f}^2)$.

\begin{figure}
    \center
    \includegraphics[scale=0.18]{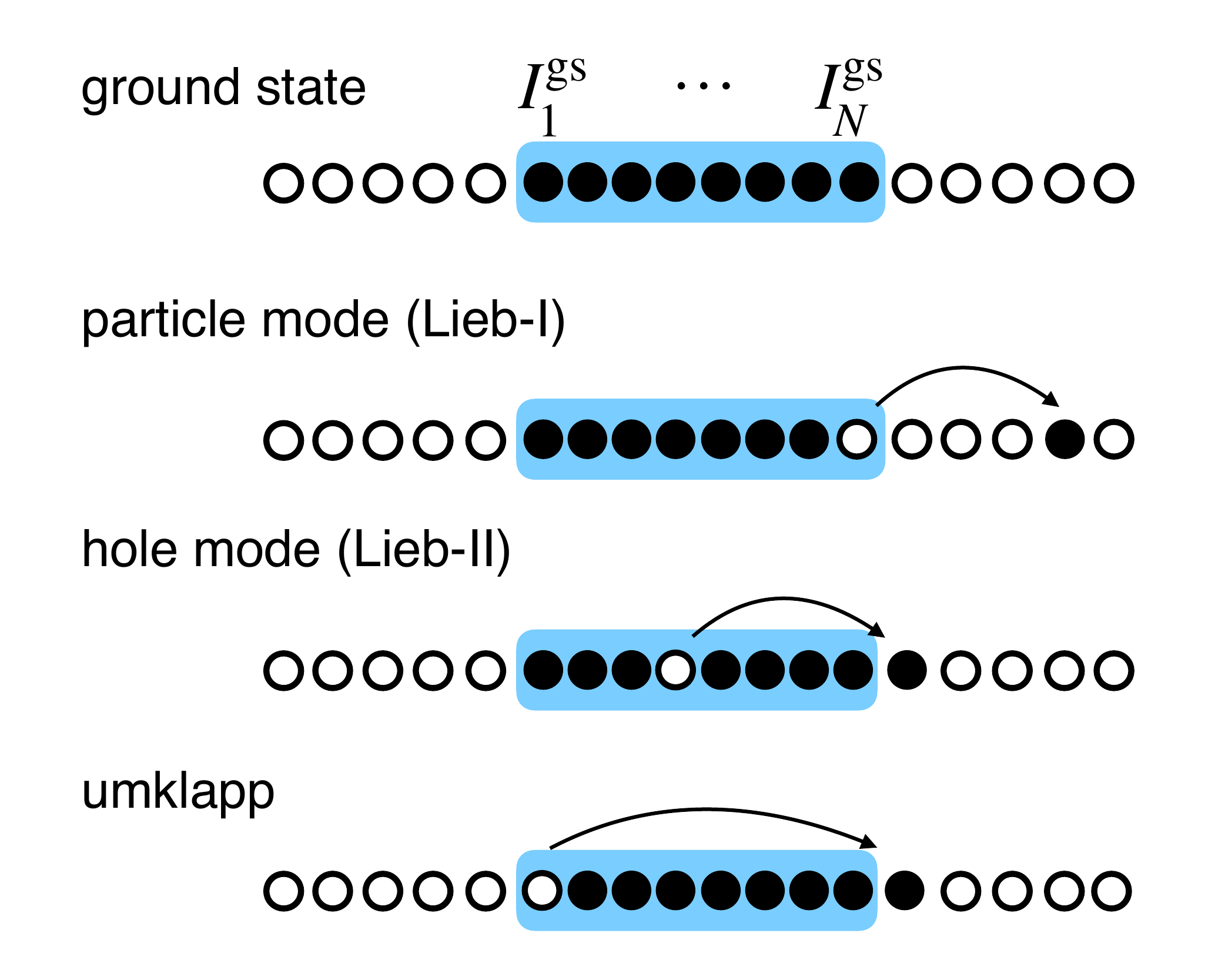}
    \caption{
    Illustration of quantum numbers $I_j$ corresponding to the ground state (top row) and the two fundamental excitations modes (middle rows) for the system of $N=9$ particles. 
    Filled (empty) points mark non-zero (zero) values.
    The umklapp excitations obtained by moving a particle from one to the other Fermi edge are illustrated in the bottom row. They can be seen as a limiting case of the hole excitation.}
    \label{fig:gs_excitations}
\end{figure}

Excited states, with the same particle number, are obtained by other choices of the quantum numbers. The two prototypical excitations are Lieb-I (particle) and Lieb-II (hole) modes, borrowing the nomenclature from the Lieb-Liniger model~\cite{1963_Lieb_PR_130_1,1963_Lieb_PR_130_2}. They correspond respectively to taking a particle from the edge of the Fermi sea and placing it away from it or by taking a particle from within the Fermi sea (thus creating a hole) and placing it right at the edge of the Fermi sea. Their dispersion relations are~\cite{PhysRevA.94.043627}
\begin{equation} \label{fundamental_modes}
    \omega_{1}(k) = \frac{2k k_F + k^2}{K}, \qquad \omega_2(k) = \frac{|2k k_F - k^2|}{K},
\end{equation}
with the Fermi momentum $k_F = \pi \rho_0$. The spectrum is gapless with the sound velocity $v_s = 2k_F/K$.

A special case of the hole mode is the umklapp state in which one particle is taken from the left edge and placed right after the right edge of the Fermi sea. This process can be repeated leading to higher umklapp states. The umklapp states have momentum $2m k_F$ with $m$ being their multiplicity and a vanishing, in the thermodynamic limit, energy. The structure of the ground states and these basic excited states is illustrated in Fig.~\ref{fig:gs_excitations}.

\section{Configuration integrals and boundary conditions}

Calculation of observables expressed in terms of the wavefunctions requires performing configuration integrals. These are of the form
\begin{equation}
	\int_{\rm L^N} {\rm d}\bfr_N f(\bfr) = \int_0^L {\rm d}r_1 \cdots \int_0^L {\rm d}r_N f(\bfr),
\end{equation}
where $\bfr = \{r_1, \dots, r_N\}$ and ${\rm d}\bfr_N$ denotes volume element of the $N$-particle configurational space and we assume that $f(\bfr)$ is a symmetric function of all the coordinates. For now we do not consider periodic boundary conditions.

The configurational space in one spatial dimension factorizes into subspaces with a specified ordering of the coordinates
\begin{align}
	&\{ (r_1, \dots, r_N) : r_i \neq r_j, r_i \in [0, L) \} = \bigcup_{\sigma \in \mathcal{P}_N} \mathcal{A}_\sigma, \\
	&\mathcal{A}_{\sigma } = \{ (r_{\sigma_1} < \dots < r_{\sigma_N}) : r_i \in [0, L) \}.
\end{align} 
In writing the configuration space we took into account that the wave function vanishes when two coordinates coincide. The symmetry of function $f(\bfr)$ implies that the contribution does not depend on the ordering of the coordinates. This allow us to write
\begin{equation}
	\int_{\rm L^N} {\rm d}\bfr_N f(\bfr) = N! \int_{\mathcal{A}_\sigma} {\rm d}\bfr_N f(\bfr).
\end{equation}
For the hard-rods, the wavefunction vanishes when two coordinates are at distance smaller than $a$. This leads to a further simplification of the configuration integral which takes an especially simple form in the hard-rods coordinates $x_j$, given by~\cite{1940705}
\begin{equation} \label{hard_rods_transformation}
	x_j = r_j - (j-1)a.
\end{equation}
Under this transformation the rods are mapped to point-like particles.
Simple manipulations give
\begin{equation}
	\int_{\rm L^N} {\rm d}\bfr_N f(\bfr) = N! \int_{\mathcal{\bar{A}}_\sigma} {\rm d}\bfx_N f(\bfx),
\end{equation}
where
\begin{equation}
	\mathcal{\bar{A}}_{\sigma } = \{ (x_{\sigma_1} < \dots < x_{\sigma_N}) : x_i \in [0, L_{\rm f}) \}
\end{equation}
At this stage we can once again use the symmetry of function $f(\bfx)$ to write
\begin{equation} \label{eq:transformation}
	\int_{\rm L^N} {\rm d}\bfr_N f(\bfr) =  \int_{L_{\rm f}^N} {\rm d}\bfx_N f(\bfx).
\end{equation}
This identity is a consequence of the symmetry of the integrand combined with the transformation~\eqref{hard_rods_transformation} and the possibility to order particles on a 1d line.

In a system with periodic boundary conditions the above discussion is modified. Consider computing normalization of the wavefunction. Here, due to a translational invariance, one coordinate can take an arbitrary value from the range $[0, L]$ in real coordinates or $[0, L_{\rm f}]$ in hard-rods coordinates. This leads to an extra overall factor $L_{\rm f}/L$ when comparing the integrals in real and hard-rods coordinates. 

This result has consequences for the normalization of the hard rods wavefunction. It implies the normalization factor in~\eqref{wave_function_F} is 
\begin{equation}
    \mathcal{N} = \frac{L}{L_{\rm f}} \times N!\, L_{\rm f}^N,
\end{equation}
where the second factor is the normalization of a wavefunction of a free fermionic theory of $N$ particles bounded to line of length $L_{\rm f}$.

\section{Form-factors}

The aim of this section is to compute the form-factors $\langle \bfm | \hat{\rho}(0)| \bfl \rangle$ of the density operator for two arbitrary eigenstates, from the knowledge of theirs wavefunctions. This matrix element denotes the probability amplitude for the system, initially in state $| \bfl \rangle$, to be promoted to state $| \bfm \rangle$ through interaction with a localized density fluctuation at the origin.

To compute the form-factors we use the wavefunction representation~\eqref{wave_function_F} and \eqref{wave_function},
\begin{equation}
	\langle \bfm | \hat{\rho}(0)| \bfl \rangle = \int {\rm d}\bfr_N \psi_{F/B}^*(\bfr| \bfm) \hat{\rho}(0) \psi_{F/B}(\bfr| \bfl).
\end{equation}
Note, the form factors do not depend on statistics as sign in \eqref{wave_function} simplifies to one.
The integrand is a symmetric function of coordinates $\bfr$. This implies that we can use the transformation~\eqref{eq:transformation} to the hard-rods coordinates which yields
\begin{equation}
	\langle \bfm | \hat{\rho}(0)| \bfl \rangle = N \int {\rm d}\bfx_N \delta(x_1) \psi_{F/B}^*(\bfx| \bfm) \psi_{F/B}(\bfx| \bfl),
\end{equation}
where we additionally used that the density operator gives $N$ identical contributions. The form-factor takes the form of a $N-1$ dimensional integral. Thanks to the Slater determinant form of the wave-function, it factorizes into a product of $1$-dimensional integrals. To this end we expand the determinants into sums over permutations, as in \eqref{wave_function_sum}. This gives
\begin{equation}
	\langle \bfm | \hat{\rho}(0)| \bfl\rangle = \frac{N}{\mathcal{N}} \sum_{\sigma, \tau \in \mathcal{P}_N} (-1)^{|\sigma| + |\tau|} \mathcal{I}(\sigma, \tau),
    \label{eq: Formfactor}
\end{equation}
where we introduced
\begin{equation}
	\mathcal{I}(\sigma, \tau) = \int {\rm d}{\bfx}_N \delta(x_1) \prod_{j=1}^N e^{i\left(\mu_{{\sigma}_j} - \lambda_{{\tau}_j} \right)x_j}.
\end{equation}
Simple computations give then
\begin{equation}
	\mathcal{I}(\sigma, \tau) = \prod_{j=2}^N \gamma(\mu_{\sigma_j}, \lambda_{\tau_j}),
\end{equation}
with
\begin{equation}
	\gamma(\mu, \lambda) = \int_0^{L_f} {\rm d}x e^{i(\mu -\lambda)x} = \frac{e^{i (\mu - \lambda){L_f}} - 1}{i(\mu - \lambda)}.
\end{equation}
The case of $\mu = \lambda$ is understood through the limiting procedure. In evaluating the integral we assumed that $\mu$ and $\lambda$ are arbitrary real parameters. However, in the formula for the form-factor, the sets of rapidities are not arbitrary. Instead they are given by the solution~\eqref{eq:Bethe} to the Bethe equation. This leads to an important simplification of the exponential term,
\begin{equation}
	\gamma(\mu_j, \lambda_m) =  \frac{e^{i a \Delta P} - 1}{i(\mu_j - \lambda_m)},
\end{equation}
where $\Delta P = P_{\bfm} - P_{\bfl}$. 
The form-factor reads now
\begin{equation}
	\langle \bfm | \hat{\rho}(0)| \bfl \rangle = \frac{N}{\mathcal{N}} \sum_{\sigma, \tau \in \mathcal{P}_N} \frac{(-1)^{|\sigma| + |\tau|}}{\gamma(\mu_{\sigma_1}, \lambda_{\tau_1})} \prod_{j=1}^N \gamma(\mu_{\sigma_j}, \lambda_{\tau_j}),
\end{equation}
where we extended the product to include $j=1$ term and in the same time we divided by it. 
The double sum over the permutations can be still simplified. By writing the product in the order dictated by  permutation $\sigma$ and changing sum over $\tau$ into a sum over $\eta = \tau \cdot \sigma^{-1}$ we obtain
\begin{equation}
	\langle \bfm | \hat{\rho}(0)| \bfl \rangle = \frac{N}{\mathcal{N}} \sum_{\sigma, \eta \in \mathcal{P}_N} \frac{(-1)^{|\eta|}}{\gamma(\mu_{\sigma_1}, \lambda_{(\eta \cdot \sigma)_1})} \prod_{j=1}^N \gamma(\mu_j, \lambda_{\eta_j}).
\end{equation}
The sum over $\sigma$ can be now readily evaluated
\begin{align}
	\frac{1}{(N-1)!}\sum_{\sigma \in \mathcal{P}_N} \frac{1}{\gamma(\mu_{\sigma_1}, \lambda_{(\eta \cdot \sigma)_1})} %&= (N-1)! \sum_{j=1}^N \frac{1}{\gamma(\mu_j, \lambda_{\eta_j})} \\
	= \frac{i \Delta P}{e^{i a \Delta P} - 1},
\end{align}
and the answer does not depend on the permutation $\eta$. The remaining sum over the permutations form a determinant of $\gamma(\mu_j, \lambda_m)$ which can be written as a Cauchy determinant. The result is 
\begin{equation} \label{form_factor}
	\langle \bfm | \hat{\rho}(0) | \bfl \rangle = \frac{\Delta P}{L}\frac{i^{N-1}}{L_{\rm f}^{N-1}} \left(1 - e^{i a \Delta P}\right)^{N-1} \mathcal{C}[\bfm, \bfl],
    %\langle \bfm | \hat{\rho}(0) | \bfl \rangle = \frac{i^{N-1}}{L_{\rm f}^N} \Delta P \left(1 - e^{i a \Delta P}\right)^{N-1} \mathcal{C}[\bfm, \bfl],
\end{equation}
with the Cauchy determinant
\begin{equation}
	\mathcal{C}[\bfm, \bfl] = \det\left(\frac{1}{\lambda_i - \mu_j} \right) = \frac{\prod_{i < j}(\lambda_j - \lambda_i)(\mu_j - \mu_i)}{\prod_{i,j}(\mu_i - \lambda_j)}.
\end{equation}
The computational complexity of the form-factor, due to the determinant form, reduces from $(N!)^2$ to $N^3$.

The form-factors, despite the simplicity of the hard-rods interactions, display properties typical to other strongly correlated systems and substantially differ from form-factors in the Tonks-Girardeau (TG) gas. In the latter case, the form-factors are 
\begin{equation} \label{form_factor_TG}
    \langle \bfm | \hat{\rho}(0) | \bfl \rangle_{\rm TG} = \begin{cases}
        \rho_0, \qquad &\bfm = \bfl, \\
        \frac{1}{L}, \qquad &\bfm/\{\mu^+\} = \bfl/\{\lambda^-\}, \\
        0, \qquad &{\rm otherwise},
    \end{cases}
\end{equation}
They are equal to $\rho_0=N/L$ when rapidities are identical; to $1/L$  when two eigenstates differ precisely by a value of a single rapidity, $\{\mu^+\}$ and $\{\lambda^-\}$, with the remaining $N-1$ rapidities exactly equal; and to zero for other rapidities. This expression can be derived from~\eqref{form_factor} by taking $a \rightarrow 0$, see in Appendix~\ref{app:TG_formfactors}. As a consequence, the form-factor in the TG gas is generically zero and is non-zero only for a very special choice of the two eigenstates.

\begin{figure}
	\includegraphics[scale=0.15]{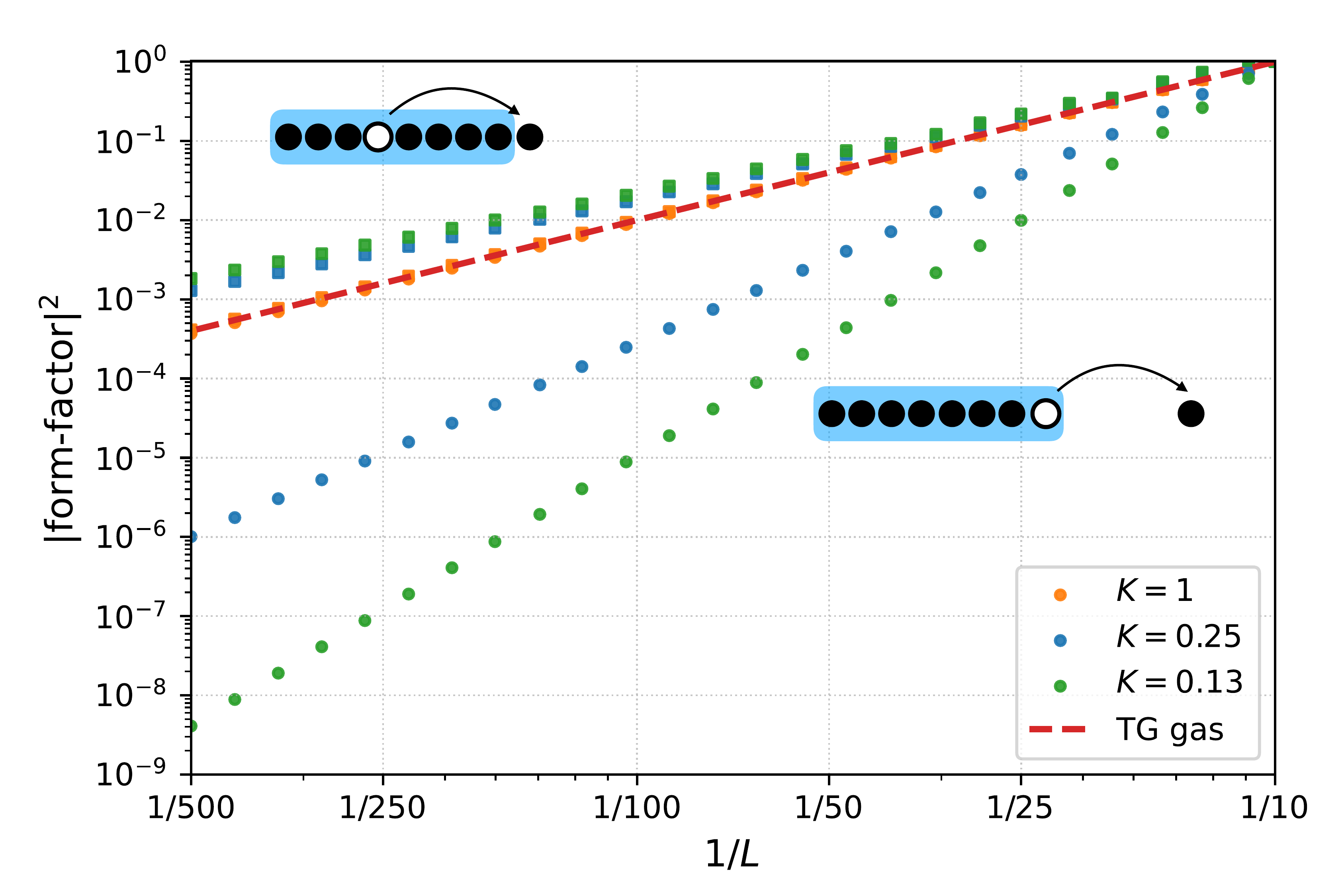}
	\caption{Form-factors between the ground state and a state containing either a single particle (circles) excitation or a single hole excitation (squares), both of fixed momentum $k_F$, as a function of the system size $L$ with density $\rho_0=1$. For $K=1$ form-factors squared scale as $1/L^2$ (shown with a black line) which is expected in a free theory. Instead for other values of $K$ the scalling is irrational with hole form-factors scaling with a smaller power while particle form-factors with a larger power.}
	\label{fig:fractional_power}
\end{figure}

This is unlike the just obtained form-factors~\eqref{form_factor}. In the quantum hard-rods the form-factors between two arbitrary eigenstates (with the same number of particles and with different momenta) are generically non-zero. Moreover, the form-factors scale with a fractional power of the system size as shown in Fig.~\ref{fig:fractional_power} for form-factors between the ground state and a state with a single elementary excitation. In the computation of the Luttinger liquid prefactors, we will see another example of this for form-factors between the ground state and umklapp states. The fractional scaling with system size is another characteristic of the strongly correlated system, with its origins rooted in the Anderson orthogonality catastrophe. Form-factors of the density operator in the quantum hard-rods share these two features with the form-factor of the Lieb-Liniger model~\cite{1990_Slavnov_TMP_82,PhysRevX.14.031048} making their dynamic response functions display similar complexities.

In the following, we will analyse further properties of the form-factors~\eqref{form_factor}. We will reveal a fine structure of them that depends on one hand on the ratio $a/L$ which for a given system is fixed and on the other hand on the value of $\Delta P$. The latter quantity depends on the choice of the two eigenstates. We start by considering the zero momentum form-factors where the properties of the form-factors do not depend on the ratio $a/L$.

\subsection{Zero momentum form-factors}

Zero momentum form-factors are form-factors evaluated between two eigenstates with the same momentum and hence $\Delta P=0$.
The density operator at zero momentum is simply a particle number operator when working in the Fock representation.
Therefore, at zero momentum, the form-factor is non-zero only for coinciding bra and ket states. 

This can be directly verified from the formula~\eqref{form_factor}. In the case $\Delta P=0$, the prefactor vanishes with $(\Delta P)^N$ and this can be tamed by the divergence of the Cauchy determinant only when all the quasi-momenta coincide. Therefore, for $\Delta P = 0$ the form-factor for different states is zero, while the diagonal form-factor is $\langle \bfl| \hat{\rho}(0)|\bfl\rangle = \rho_0$ just like in the TG case.

\subsection{Non-zero momentum form-factors}

We assume that the two eigenstates have different momenta and $\Delta P \neq 0$. Still, the prefactor in~\eqref{form_factor} can vanish if $a \Delta P$ is a multiple of $2\pi$. Since $\Delta P$ is a multiple of $2\pi/L$ according to~(\ref{eq:defP}), the necessary, but not sufficient, condition for the prefactor to vanish is that the ratio $a/L$ is a rational number. This condition depends on the system Hamiltonian (\ref{eq:ham}). 
If the ratio $a/L$ is irrational, then the prefactor~\eqref{form_factor} never vanishes. 
We also note that in such a case, 
the quasi-momenta between the two eigenstates can never coincide, 
and the form-factor is always finite. This is a generic case. The non-generic case we discuss in the next section.

\subsection{Rational case - TG alike}

When $a/L$ is rational, there are states for which $a \Delta P$ is a multiple of $2\pi$. For such $\Delta P$, the prefactor of the form-factor vanishes. At the same time, some of the quasi-momenta can now coincide, which makes the Cauchy determinant diverge. This interplay results in non-zero form-factors only when either $N$ or $N-1$ quasi-momenta coincide. In the former case, we obtain the diagonal form-factor discussed above.  Instead, for one pair of differing quasi-momenta, the form-factor is constant and equal to $1/L$.

The structure of these class of form-factors is exactly like in the Tonks-Girardeau gas, see eq.~\eqref{form_factor_TG}. The crucial difference is that whereas in the Tonks-Girardeau gas the form-factor at finite momentum is non-zero only when $N-1$ quantum numbers coincide, in the hard-rod system this selection rule applies only to situations when $a\Delta P$ is a multiple of $2\pi$. We will comment on a physical relevance of this finding when discussing the DSF to which we turn now.

\section{Dynamic structure factor}

\begin{figure*}[]
	\includegraphics[width=\linewidth]{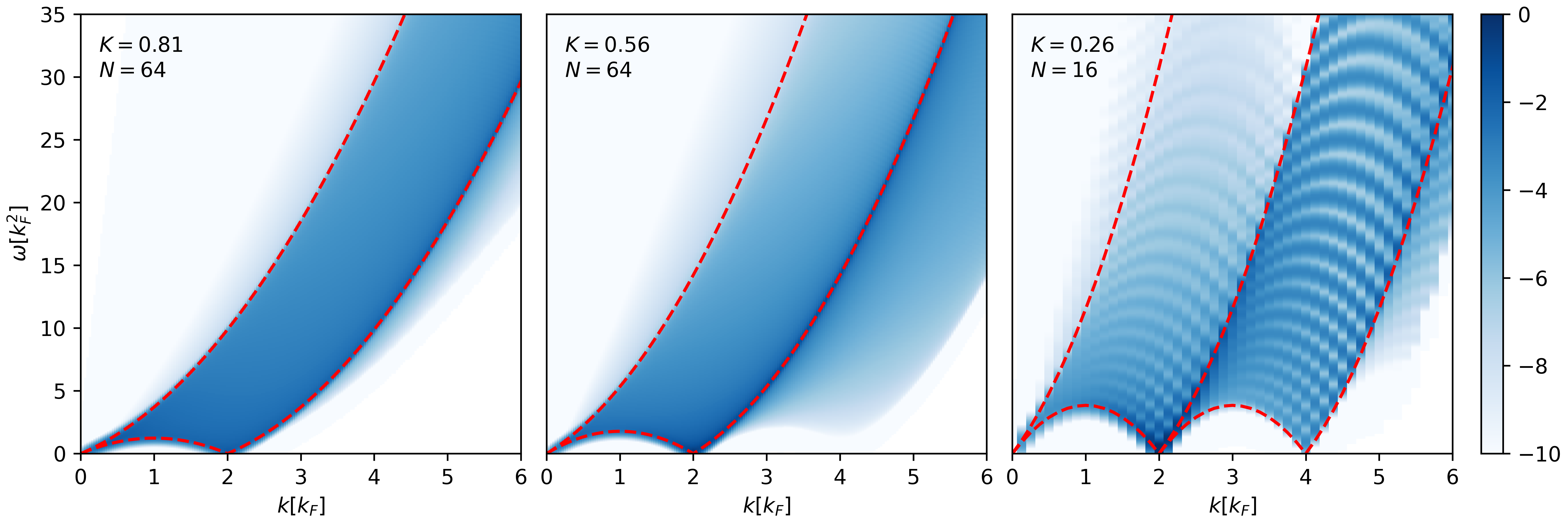}
	\caption{
    Color maps of the dynamic structure factor, presented on a logarithmic scale as ${\rm log}[S(k,\omega)]$ are shown for three representative values of the Luttinger parameter $K$, with $\rho_0=1$ and for finite number $N$ of particles as indicated in the plots.
    The red dashed lines indicate the dispersion relations of the two fundamental modes~\eqref{fundamental_modes}, and for $K=0.26$ the dispersion relation of the second hole-mode is also shown.
    For $K=0.26$, finite-size effects manifest as stripes, also visible in the fixed-momentum cuts of Fig~\ref{fig:dsf_fixed}; incomplete summation of high-$k$ excitations, despite including over $10^9$ states (see~Tab.~\ref{tab:states}), produces bright patches.
    }
    \label{fig:dsf}
\end{figure*}

The dynamic structure factor (DSF) in the spectral representation is defined as the Fourier transform of the real-space, real-time density-density correlation function,
\begin{equation} \label{spectral_sum}
	S(k, \omega) = 2\pi L\sum_{\bfm} |\langle {\bfm} | \hat{\rho}(0) | {\rm gs} \rangle|^2 \delta(\omega - (E_{\bfm} - E_{\rm gs})) \delta_{k, P_{\bfm}},
\end{equation}
with form-factors $\langle \bfm | \hat{\rho}(0) | {\rm gs} \rangle$ of the density operator between the ground state and an eigenstate of energy $E_{\bfm}$. The sum is over all allowed sets of quantum numbers.
The DSF provides information about how density fluctuations propagate through the system and how the system responds to external perturbations that transfer momentum $k$ and energy $\omega$.
A contribution to the DSF at a given energy $\omega$ only comes from transitions where the energy difference between the excited state $|{\bfm}\rangle$ (with energy $E_{\bfm}$) and the ground state $|{\rm gs} \rangle$ (with energy $E_{\rm gs}$) is exactly equal to $\omega$.
It also implies that only transitions where the momentum $P_{\bfm}$ of the excited state $|{\bfm}\rangle$ matches the momentum transfer $k$ contribute to the DSF (knowing the ground state $|{\rm gs}\rangle$ has zero momentum, $P_{\rm gs}=0$).
The DSF probes only those quantum transitions that conserve both energy and momentum, thereby mapping out the permissible excitation spectrum in the energy-momentum space of the system.

The computation of the dynamic structure factor reduces to the evaluation of the spectral sum~\eqref{spectral_sum} when equipped with the analytical expressions for form-factors~(\ref{form_factor}).
This can be effectively implemented on a computer in a similar fashion as was done for other Bethe ansatz solvable models~\cite{PhysRevLett.95.077201,2006_Caux_PRA_74,1742-5468-2007-01-P01008,2009_Caux_JMP_50,Li_2023}. The idea is to organise the spectral sum into number of particle-hole excitations and introduce a cut-off $I_{\rm max}$ for the maximal (absolute) value of the quantum numbers, $|I_j| \leq I_{\rm max}$ and for the number of particle-hole excitations. The spectral sum is thus restricted to a finite, however large, number of elements. The numerical results presented here were obtained using the \textsc{ABACUS} algorithm~\cite{ABACUS} augmented by us with the functionality to handle hard-rods systems. The computations were performed on a standard laptop. \textsc{ABACUS} applies heuristics to optimally scan the Hilbert space, using the absolute values of the form factors to determine the ordering.

The accuracy of the evaluation can be tested with the $f$-sum rule
\begin{equation} \label{fsum_rule}
	\int \frac{{\rm d}\omega}{2\pi} \omega S(k, \omega) = \rho_0 k^2.
\end{equation}
The levels of saturation of the $f$-sum rule, defined as the ratio of the left and right hand sides of \eqref{fsum_rule}, for the numerical data presented below are shown in Table~\ref{tab:fsum_rule}. The computations are stopped when the f-sum rule acquires the desired saturation. This together with the heuristics of the ABACUS algorithm renders the cut-off quantum number in practice immaterial as, if set large enough, is never reached in the scanning procedure. 

For reference, we also quote the exact result for the correlation function in the Tonks-Girardeau gas
\begin{equation}
    S_{\rm TG}(k, \omega) = \frac{\Theta(\omega - \omega_2^{\rm TG}(k)) - \Theta(\omega - \omega_1^{\rm TG}(k))}{2k},
\end{equation}
where $\Theta(x)$ is the Heaviside $\Theta$-function and $\omega_{1,2}^{\rm TG}$ are the dispersions~\eqref{fundamental_modes} of the two fundamental modes for the Tonks-Girardeau gas, namely with $K=1$. The characteristic feature of the DSF in the TG gas is its constant lineshape in $\omega$ between the two limiting dispersion lines and for fixed $k$.

\begin{table}[h]
\centering
\begin{tabular}{c|ccc}
\toprule
$k$  & $K=0.81$ & $K=0.56$ & $K=0.26$ \\
 & $N=64$ & $N=64$ & $N=16$ \\
\midrule
$k = k_F$ & $ 0.9998$ & $0.9998$ & $0.9999$ \\
$k= 2k_F$  & $0.9994$ & $0.9990$ & $0.9999$ \\
$k= 4k_F$ & $0.9945$ & $ 0.9904$ & $0.9999$ \\
$k= 6 k_F$ & $0.9827$ & $0.9698$ & $0.9855$ \\
\bottomrule
\end{tabular}
\caption{Levels of saturation of the $f$-sum rule~\eqref{fsum_rule} for different systems. The density is $\rho_0 = 1$ for all cases.} 
\label{tab:fsum_rule}
\end{table}%

In Fig.~\ref{fig:dsf} we show color heatmaps of the dynamic structure factor $S(k, \omega)$ for three representative values of the Luttinger liquid parameters $K$. With decreasing value of $K$ (or increasing rods length $a$ with the density $\rho_0$ fixed) we observe a significant widening of the response of the system, showing departure from the free fermionic behavior and entering a strongly correlated regime. Importantly the 
signal shifts towards higher momenta, for $K=0.26$, its actually stronger in the second band of excitations than in the fundamental one given by the two Lieb's modes. 

\begin{figure*}[]
	\includegraphics[width=\linewidth]{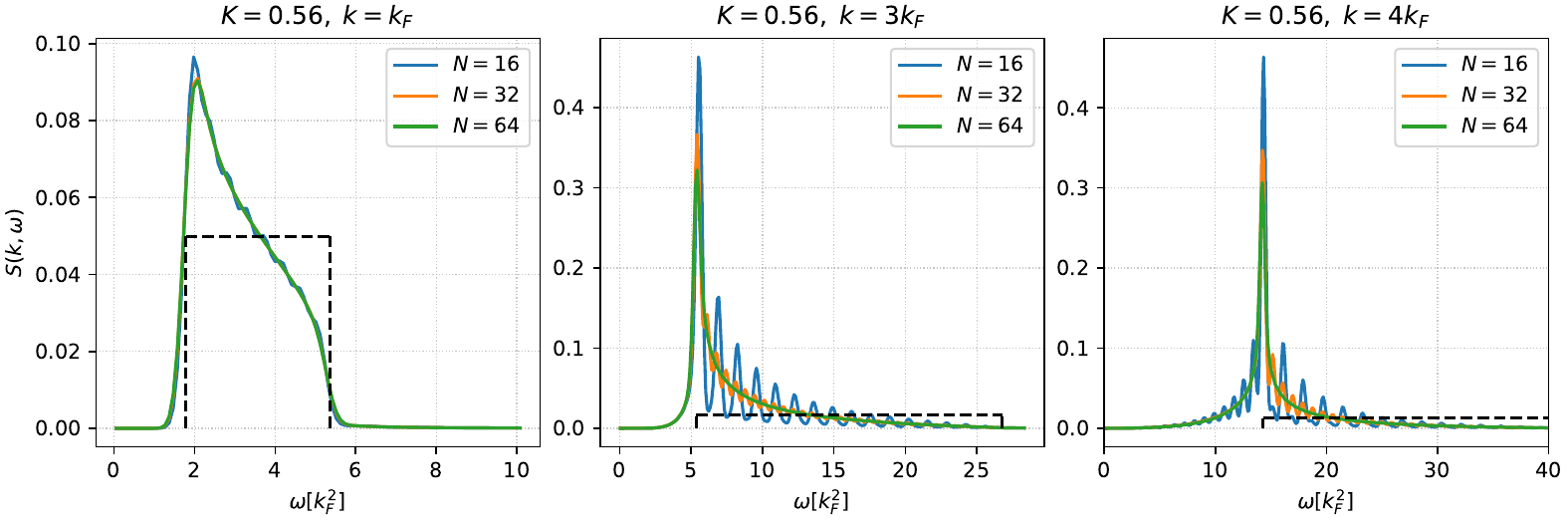}
	\caption{
    Fixed-momentum cuts of the dynamic structure factor $S(k,\omega)$ are shown for various $k$ and system sizes $N$.
    The lineshapes are characterized by a peak in the vicinity of the lower threshold and a long high-energy tail.
    We observe that finite-size effects for $N=64$ are essentially below the resolution of the figure, except in the vicinity of the peak, where such effects are expected to be more pronounced. 
    The dashed black lines is $S_{\rm mTG}(k,\omega)$, the DSF of the modified Tonks-Girardeau gas, see Eq.~\eqref{DSF_guess}.}
    \label{fig:dsf_fixed}
\end{figure*}

To display a finer structure of $S(k,\omega)$, in Fig.~\ref{fig:dsf_fixed} we present its fixed momentum cuts. The resulting lineshape differs significantly from the free fermionic shape and exhibits singularities along the lower threshold of excitations in agreement with the predictions of the non-linear Luttinger liquid theory~\cite{Imambekov_2008,Imambekov_2009} with $K<1$. 

To further exhibit a strongly correlated nature of the system, we confront these results with a simple guess for the DSF
\begin{equation} \label{DSF_guess}
    S_{\rm mTG}(k, \omega) = K\frac{\Theta(\omega - \omega_2(k)) - \Theta(\omega - \omega_1(k))}{2k},
\end{equation}
which inherits from the TG gas the constant profile in energy, but uses the exact dispersion lines. The proportionality factor is chosen such that the $f$-sum rule holds. This comparison highlights the complex behavior of quantum hard rods, manifested in the non-trivial effects of rod dimensionality on the excitation spectrum. It underscores how the model deviates from simple free-fermion-like behavior.

The strongly correlated nature of the system can also be observed in the number of relevant states contributing to the DSF. This is shown in Table~\ref{tab:states} and, besides physical importance, has also practical implications: calculations for systems with tightly packed particles become computationally very demanding when system sizes increase.

\begin{table}[h]
\centering
\begin{tabular}{c|ccc}
\toprule
$N$ & $K =  0.81$ & $K = 0.56$ & $K = 0.26$ \\
\midrule
8 &  $71\,805$ & $132\,134$ & $4\,058\,591$ \\
16  & $560\,836$ & $1\,443\,177$ & $1\,016\,744\,377$  \\
32 & $27\,177\,166$ & $47\,303\,853$ & --- \\
64  & $113\,311\,903$ & $451\,838\,123$ & ---  \\
\bottomrule
\end{tabular}
\caption{Number of states needed to reach at least $99\%$ saturation of the $f$-sum rule for different Luttinger liquid parameters $K$ and different system sizes $N$. These numbers illustrate the quickly increasing complexity of the structure behind $S(k,\omega)$ with decreasing value of $K$. }
\label{tab:states}
\end{table}%

\begin{figure}
	\center
	\includegraphics[scale=0.55]{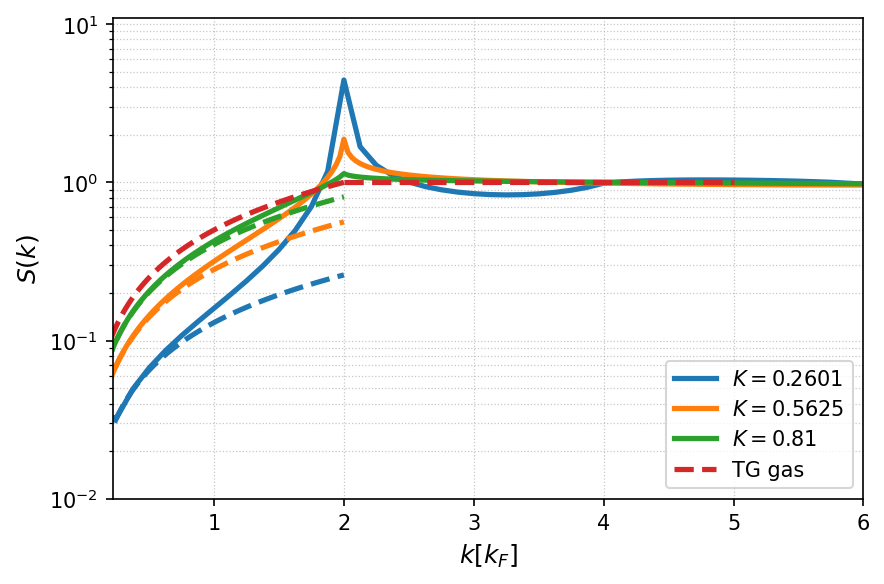}   
	\caption{
    Static structure factor $S(k)$ of the system with $N=16$ and representative values of $K$ as indicated in legend. 
    The corresponding in color dashed lines represent asymptotic behavior $S(k\sim0) \approx |k|/v_s$ known from TL theory.
    The only red dashed line represents known TG gas behavior when $K\to 1$.
    }
	\label{fig:static}
\end{figure}

Finally, in Fig.~\ref{fig:static}, we present results for the static structure factor 
\begin{equation}
    S(k)=\int \frac{{\rm d}\omega}{2\pi} S(k, \omega),
\end{equation}
marked by solid lines and together with its universal behavior $S(k) \sim |k|/v_s$ near $k \sim 0$ for gapless 1d systems (dashed lines), as well as the corresponding behavior for the Tonks--Girardeau gas. The interactions present themselves in a number of effects. 
The characteristic decay at zero momentum reflects the strong correlations present in the system, whereas the saturation to a value of density (here $\rho_0 = 1$) at higher momenta is indicative of an uncorrelated regime.
The most prominent feature is the appearance of a peak at $k = 2k_F$, whose height, along with the depth of the subsequent dip, increases with decreasing $K$. 
These findings are consistent with previous numerical results reported in the literature~\cite{PhysRevA.94.043627} and with the expected charge density wave phase of the Luttinger liquid with $K<1$. Peaks of the same nature appear also in the response function in a metastable super Tonks-Girardeau gas~\cite{PhysRevLett.110.125302}. For smaller values of $K$ additional peaks are expected to emerge at multiplicities of $2k_F$.These peaks set the preference in the spacing between the hard rods in the physical space. The regime of large momenta and small $K$ is difficult to access numerically as many excited states contribute to the spectral sum. To overcome these difficulties in the next section we turn to analytical methods based on the Luttinger liquid theory and extensions thereof.

\section{Real-space correlation and Luttinger liquid}

The quantum hard-rods model belongs to the universality class of Tomonaga-Luttinger (TL) liquids describing universal low-energy physics of gapless one-dimensional quantum fluids~\cite{GiamarchiBOOK}. The central quantity characterizing a TL liquid is the Luttinger parameter $K$ introduced above. For further convenience we recall here the relation between it and the rods-length
\begin{equation} \label{LL_K}
 	K = \left(1 - \rho_0 a\right)^2.
\end{equation}

In the following, we analyze the real-space correlator, defined as
\begin{equation}
    S(r) = \rho_0^2 + \int \frac{{\rm d}k}{2\pi} e^{ikr} (S(k) - \rho_0)
\end{equation}
and relate it to the prediction of the TL theory, in particular, to the asymptotic form
\begin{equation} \label{LL_S(x)}
	\frac{S(r)}{\rho_0^2} \approx 1 - \frac{K}{2(k_F r)^2} + \sum_{m \geq 1} \frac{A_m \cos(2mk_F r)}{(\rho_0 r)^{2m^2 K}}.
\end{equation}
of the equal-time correlator at distances $r \gg \rho_0^{-1}$. Here, $A_m$ are non-universal constants that depend on the specifics of the microscopic model. 

\subsection{Prefactors of the static correlation function}

The prefactors $A_m$ in the Luttinger liquid formula~\eqref{LL_S(x)} for $S(r)$ can be computed from the form-factors~\cite{2011_Shashi_PRB_84,Shashi_2012}. They are given by a specific scaling of the form-factor between the ground state and $m$-umklapp state
\begin{equation} \label{A_m_def}
	A_m = \lim_{\rm th} \frac{2}{\rho_0^{2}} \left(\frac{\rho_0 L}{2\pi}\right)^{2m^2 K} |\langle m, N|\hat{\rho}(0)|N\rangle|^2.
\end{equation}
The thermodynamic limit is taken by sending $N, L \rightarrow \infty$ with $\rho_0 = N/L$ fixed. $|N\rangle$ and $|N, m\rangle$ denote the $N$-particle ground state and $N$-particle state with $m$-umklapps respectively. The quasi-momenta $\bfm$ of the latter state, with respect to their ground state counterparts $\bfl$, are
\begin{equation}
	\mu_j^{(m)} = \lambda_j + \frac{2\pi}{L_f} \sqrt{K} m,
\end{equation}
where we used the Luttinger liquid parameter $K$, defined in Eq.~\eqref{LL_K} The momentum of the $m$-umklapp state is $\Delta P = 2k_F m$.

\begin{figure}
	\center
	\includegraphics[scale=0.48]{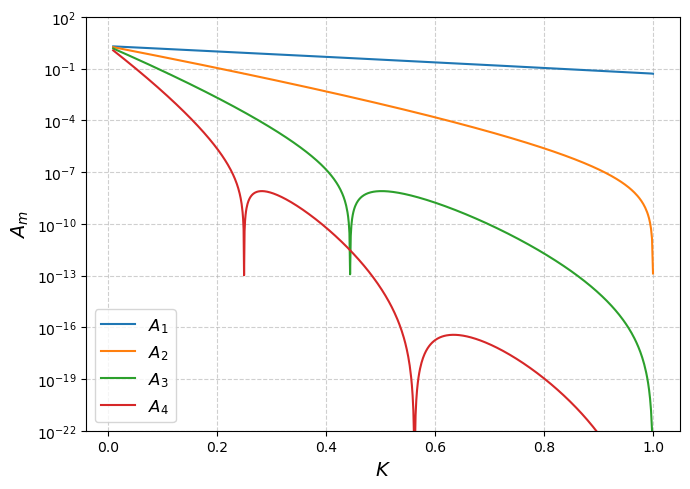}
	\caption{The prefactors $A_m$ of the density correlation function $S(x)$ as a function of the  Luttinger parameter $K$, for $m=1,2,3,4$ (from top to bottom). In the free fermionic limit $K\rightarrow 1$, only $A_1$ is non-zero; $A_3$ vanishes for $K=1$ and $K=4/9$, while $A_4$ vanishes at three distinct values of $K=1,9/16, 1/4$. 
    In general vanishing of $A_m$ implies that the system is in the TG alike regime.}
	\label{fig:prefactors}
\end{figure}

The form-factor between the ground state and the umklapp state can be expressed in terms of the Barnes G functions. The computations presented in Appendix~\ref{app:prefactor} lead to the following exact result
\begin{align}
	|\langle m, N|\hat{\rho}(0)|N\rangle| = \frac{k_F m \sqrt{K}}{\sin (a k_F m)} G(1+\nu)G(1-\nu) \nonumber \\ \times \frac{G^2(N+1)}{G(N+1+\nu)G(N+1-\nu)},
\end{align}
where $\nu = m \sqrt{K}$.
The thermodynamic limit from~\eqref{A_m_def} can now be readily taken. We use asymptotic ($N\to  \infty$) behavior,
\begin{equation}
    \frac{G^2(N+1)}{G(N+1+\nu)G(N+1-\nu)} =  N^{\nu^2} \times \left( 1 + \mathcal{O}(1/N^2)\right),
\end{equation}
and we find
\begin{equation} \label{A_m_result}
	A_m = \frac{2K(k_F m)^2 (2\pi)^{-2m^2 K}}{\rho_0^{2} \sin^2(a k_F m)}  G^2(1 + m\sqrt{K}) G^2(1 - m \sqrt{K}). 
\end{equation}
In Fig.~\ref{fig:prefactors} we plot $A_m$ as a function of $K$ with $a$ determined from the relation~\eqref{LL_K}. The analytical expression for $K \rightarrow 1$ reproduces density–density fluctuations of a free-fermion theory where $A_1 = 1/(2\pi^2)$ while the prefactors for higher umklapp terms vanish. 

The coefficients $A_m$ with $m \geq 2$ vanish also when $\rho_0 a = n/m$ for $0 \leq n \leq m-2$. This is because the Barnes function is zero at zero and at negative integers. In fact $G(-k + x) \sim x^{k+1}$ for small $x$ and $k=0, 1, 2, \dots$. 
$A_1$ term is always non-zero because the vanishing of the Barnes function is exactly compensated by the vanishing of the sine term in the denominator. 
The prefactor $A_2$ vanishes for $\rho_0 a = 0$, while the prefactor $A_3$ vanishes for $\rho_0 a = 0$ and also $\rho_0 a = 1/3$ corresponding to $K = 1$ and $K = 4/9$, respectively. 
Higher prefactors vanish in an increasing number of points. 
The variation of these prefactors is shown in Fig.~\eqref{fig:prefactors}. The vanishing of the $A_m$ coefficients requires that $\rho_0 a$ is rational, which directly corresponds to the condition for TG-like behavior, as described in Section IV. The prefactors reach their maximal, $m$-independent value, $2$ for $K=0$. Approximating the prefactors in~\eqref{LL_S(x)} by this maximal value we find
\begin{equation} \label{S(x)_TP}
    \frac{S(r)}{\rho_0^2} \approx \vartheta_3\left(r k_F, (r \rho_0)^{-2K}\right). % - \frac{K}{2(k_F r)^2},
\end{equation}
with $\vartheta_3(u, q)$ the Jacobi theta function. This provides a rough approximation for the full TL prediction~\eqref{LL_S(x)} in a tightly packed regime.

Finally, the formula~\eqref{A_m_def} for prefactors $A_m$ is a nontrivial test of the expression for the form-factors. Specifically, it requires that the scaling of the form-factors with the system size $L$ is irrational, which is generally expected in interacting theories and is reminiscent of the Anderson orthogonality catastrophe. The analytic result~\eqref{A_m_result} shows that the precise form of this scaling, for the umklapp form-factors, is in agreement with the predictions of the Luttinger liquid theory.

\subsection{Numerical results}

\begin{figure}
	\center
	\includegraphics[scale=0.48]{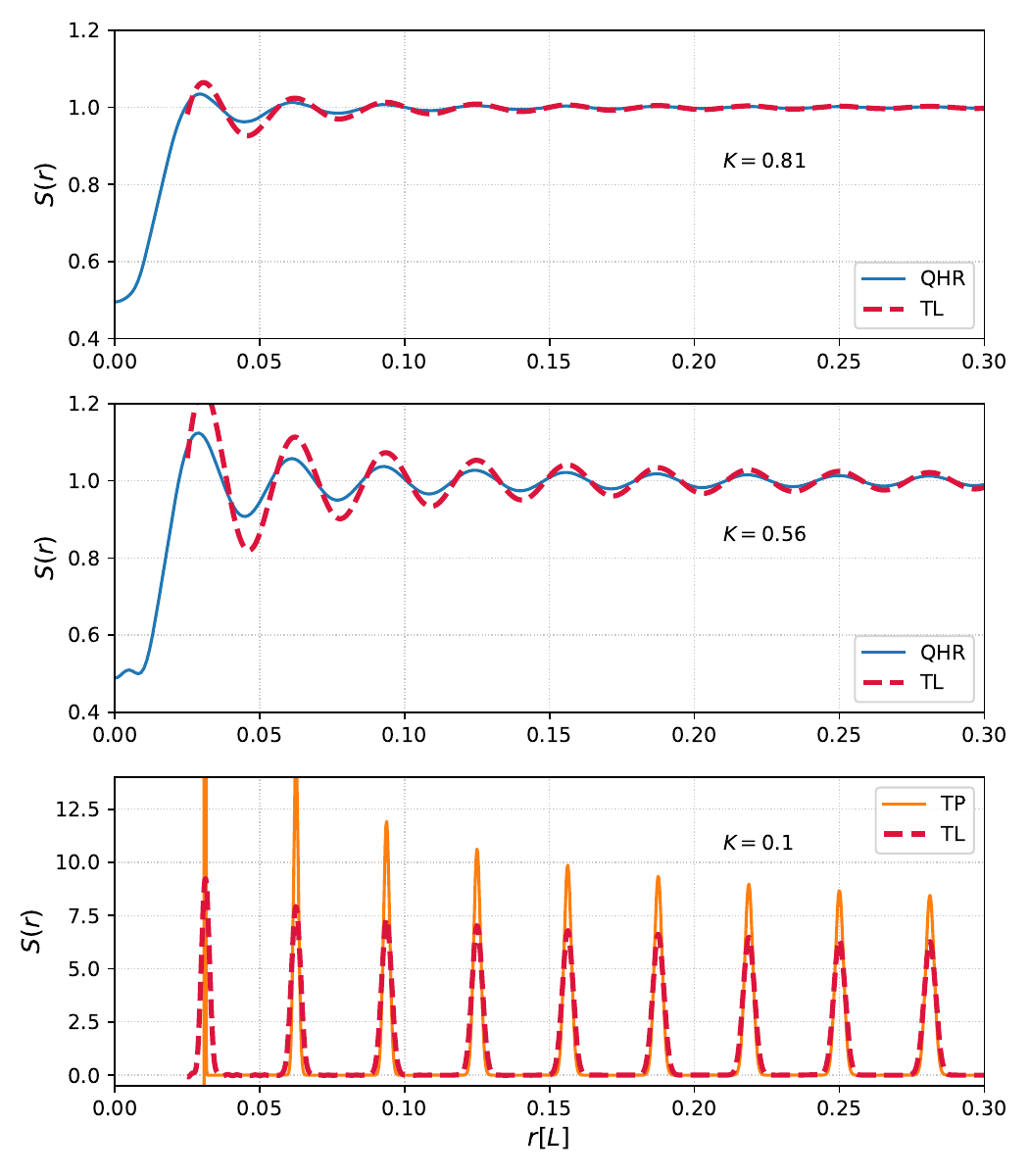}
	\caption{Equal-time correlation function $S(r)$. In the first two panels we compare results of ABACUS (QHR) for $N=32$ particles with TL theory. In the bottom panel, we compare the TL theory with a further approximation~\eqref{S(x)_TP} valid in the tightly packed (TP) regime of small $K$. The correlation function reveals a quasi-crystal structure of the gas emerging upon decreasing $K$ (or increasing $\rho_0 a$), while the comparison cross-validates the presented results.}
	\label{fig:S(x)}
\end{figure}

In Fig.~\ref{fig:S(x)}, we present a comparison between results of ABACUS and the predictions of Luttinger liquid theory for $S(r)$. We also present predictions of Luttinger liquid theory for small values of $K$ where numerical evaluation of the DSF is hard.

For \(K = 0.81\) and \(K = 0.56\), we observe the convergence between the results from the microscopic model of quantum hard-rods and predictions of the effective theory in the relevant regime of $x \gg \rho_0$.  
The prediction for the densely packed ($K\to0)$ regime, shown in the lower panel, illustrates the behavior of a strongly interacting hard rods system. In this case, exactly $N$ peaks emerge within the interval $[0, L]$. This reflects the expected physical behavior of a densely packed hard-rod system: fixing the position of one particle at $x = 0$ constrains the position of the $n$-th particle to a narrow region centered around $n a$, where $a$ denotes the rod length. In momentum space it would manifest itself in appearance of additional peaks in $S(k)$ mentioned in Section V.

\section{Conclusions}

In summary, we have shown that the one-dimensional quantum hard-rods model, despite its apparent simplicity, exhibits rich and nontrivial response properties characteristic of strongly correlated systems. 
By deriving closed-form expressions for density form factors and introducing a original semi-analytical framework for structure factors, we have provided accurate and efficient tools for studying its static and dynamic behavior. 
The results confirm Tomonaga–Luttinger liquid universality and Tonks–Girardeau–like behavior in the specific regime of parameters, and clarify the scaling properties of form factors in interacting systems.

An interesting open question is whether the TG alike behavior of the form-factors is visible in the dynamic structure factor for thermodynamically large systems. We relegate answering this question to future works. Another interesting directions are the connection with hard-rod deformations of the XXZ spin chains~\cite{PhysRevE.104.044106,PhysRevE.104.064124} and the $\bar{T}T$-deformation, which at semiclassical level changes the width of the particles~\cite{Cardy:2022aa}.

Our findings establish quantum hard rods as a reliable benchmark for testing theories of quantum many-body physics, offering stringent reference points for numerical simulations and guiding future analytical work.

\begin{acknowledgments}
We acknowledge Jean-Sébastien Caux for developing the ABACUS library and making it publicly available. 

M.P.~acknowledges support from the National Science Centre, Poland, under the OPUS grant 2022/47/B/ST2/03334.
\end{acknowledgments}

\section*{Data availability statement}

The data that support the findings of this article are openly available at~\cite{dataQHR}.

\bibliography{literature}

\begin{thebibliography}{47}%
\makeatletter
\providecommand \@ifxundefined [1]{%
 \@ifx{#1\undefined}
}%
\providecommand \@ifnum [1]{%
 \ifnum #1\expandafter \@firstoftwo
 \else \expandafter \@secondoftwo
 \fi
}%
\providecommand \@ifx [1]{%
 \ifx #1\expandafter \@firstoftwo
 \else \expandafter \@secondoftwo
 \fi
}%
\providecommand \natexlab [1]{#1}%
\providecommand \enquote  [1]{``#1''}%
\providecommand \bibnamefont  [1]{#1}%
\providecommand \bibfnamefont [1]{#1}%
\providecommand \citenamefont [1]{#1}%
\providecommand \href@noop [0]{\@secondoftwo}%
\providecommand \href [0]{\begingroup \@sanitize@url \@href}%
\providecommand \@href[1]{\@@startlink{#1}\@@href}%
\providecommand \@@href[1]{\endgroup#1\@@endlink}%
\providecommand \@sanitize@url [0]{\catcode `\\12\catcode `\$12\catcode
  `\&12\catcode `\#12\catcode `\^12\catcode `\_12\catcode `\%12\relax}%
\providecommand \@@startlink[1]{}%
\providecommand \@@endlink[0]{}%
\providecommand \url  [0]{\begingroup\@sanitize@url \@url }%
\providecommand \@url [1]{\endgroup\@href {#1}{\urlprefix }}%
\providecommand \urlprefix  [0]{URL }%
\providecommand \Eprint [0]{\href }%
\providecommand \doibase [0]{https://doi.org/}%
\providecommand \selectlanguage [0]{\@gobble}%
\providecommand \bibinfo  [0]{\@secondoftwo}%
\providecommand \bibfield  [0]{\@secondoftwo}%
\providecommand \translation [1]{[#1]}%
\providecommand \BibitemOpen [0]{}%
\providecommand \bibitemStop [0]{}%
\providecommand \bibitemNoStop [0]{.\EOS\space}%
\providecommand \EOS [0]{\spacefactor3000\relax}%
\providecommand \BibitemShut  [1]{\csname bibitem#1\endcsname}%
\let\auto@bib@innerbib\@empty
%</preamble>
\bibitem [{\citenamefont {Quintanilla}\ and\ \citenamefont
  {Hooley}(2009)}]{Quintanilla_2009}%
  \BibitemOpen
  \bibfield  {author} {\bibinfo {author} {\bibfnamefont {J.}~\bibnamefont
  {Quintanilla}}\ and\ \bibinfo {author} {\bibfnamefont {C.}~\bibnamefont
  {Hooley}},\ }\bibfield  {title} {\bibinfo {title} {The strong-correlations
  puzzle},\ }\href {https://doi.org/10.1088/2058-7058/22/06/38} {\bibfield
  {journal} {\bibinfo  {journal} {Physics World}\ }\textbf {\bibinfo {volume}
  {22}},\ \bibinfo {pages} {32} (\bibinfo {year} {2009})}\BibitemShut {NoStop}%
\bibitem [{\citenamefont {Giamarchi}(2004{\natexlab{a}})}]{Giamarchi2004}%
  \BibitemOpen
  \bibfield  {author} {\bibinfo {author} {\bibfnamefont {T.}~\bibnamefont
  {Giamarchi}},\ }\href@noop {} {\emph {\bibinfo {title} {Quantum Physics in
  One Dimension}}}\ (\bibinfo  {publisher} {Oxford University Press},\ \bibinfo
  {address} {Oxford},\ \bibinfo {year} {2004})\BibitemShut {NoStop}%
\bibitem [{\citenamefont {Franchini}(2017)}]{Franchini2016}%
  \BibitemOpen
  \bibfield  {author} {\bibinfo {author} {\bibfnamefont {F.}~\bibnamefont
  {Franchini}},\ }\href
  {https://doi.org/https://doi.org/10.1007/978-3-319-48487-7} {\emph {\bibinfo
  {title} {An Introduction to Integrable Techniques for One-Dimensional Quantum
  Systems}}}\ (\bibinfo  {publisher} {Springer},\ \bibinfo {address} {Cham,
  Switzerland},\ \bibinfo {year} {2017})\ \bibinfo {note} {lecture Notes in
  Physics, Vol. 940}\BibitemShut {NoStop}%
\bibitem [{\citenamefont {Lieb}(1963)}]{1963_Lieb_PR_130_2}%
  \BibitemOpen
  \bibfield  {author} {\bibinfo {author} {\bibfnamefont {E.~H.}\ \bibnamefont
  {Lieb}},\ }\bibfield  {title} {\bibinfo {title} {{Exact Analysis of an
  Interacting Bose Gas. II. The Excitation Spectrum}},\ }\href
  {https://doi.org/10.1103/PhysRev.130.1616} {\bibfield  {journal} {\bibinfo
  {journal} {Phys. Rev.}\ }\textbf {\bibinfo {volume} {130}},\ \bibinfo {pages}
  {1616} (\bibinfo {year} {1963})}\BibitemShut {NoStop}%
\bibitem [{\citenamefont {Paredes}\ \emph {et~al.}(2004)\citenamefont
  {Paredes}, \citenamefont {Widera}, \citenamefont {Murg}, \citenamefont
  {Mandel}, \citenamefont {F{\"o}lling}, \citenamefont {Cirac}, \citenamefont
  {Shlyapnikov}, \citenamefont {H{\"a}nsch},\ and\ \citenamefont
  {Bloch}}]{Paredes2004}%
  \BibitemOpen
  \bibfield  {author} {\bibinfo {author} {\bibfnamefont {B.}~\bibnamefont
  {Paredes}}, \bibinfo {author} {\bibfnamefont {A.}~\bibnamefont {Widera}},
  \bibinfo {author} {\bibfnamefont {V.}~\bibnamefont {Murg}}, \bibinfo {author}
  {\bibfnamefont {O.}~\bibnamefont {Mandel}}, \bibinfo {author} {\bibfnamefont
  {S.}~\bibnamefont {F{\"o}lling}}, \bibinfo {author} {\bibfnamefont
  {I.}~\bibnamefont {Cirac}}, \bibinfo {author} {\bibfnamefont {G.~V.}\
  \bibnamefont {Shlyapnikov}}, \bibinfo {author} {\bibfnamefont {T.~W.}\
  \bibnamefont {H{\"a}nsch}},\ and\ \bibinfo {author} {\bibfnamefont
  {I.}~\bibnamefont {Bloch}},\ }\bibfield  {title} {\bibinfo {title}
  {Tonks--girardeau gas of ultracold atoms in an optical lattice},\ }\href
  {https://doi.org/10.1038/nature02530} {\bibfield  {journal} {\bibinfo
  {journal} {Nature}\ }\textbf {\bibinfo {volume} {429}},\ \bibinfo {pages}
  {277} (\bibinfo {year} {2004})}\BibitemShut {NoStop}%
\bibitem [{\citenamefont {Kinoshita}\ \emph {et~al.}(2004)\citenamefont
  {Kinoshita}, \citenamefont {Wenger},\ and\ \citenamefont
  {Weiss}}]{Kinoshita2004}%
  \BibitemOpen
  \bibfield  {author} {\bibinfo {author} {\bibfnamefont {T.}~\bibnamefont
  {Kinoshita}}, \bibinfo {author} {\bibfnamefont {T.}~\bibnamefont {Wenger}},\
  and\ \bibinfo {author} {\bibfnamefont {D.~S.}\ \bibnamefont {Weiss}},\
  }\bibfield  {title} {\bibinfo {title} {Observation of a one-dimensional
  tonks-girardeau gas},\ }\href {https://doi.org/10.1126/science.1100700}
  {\bibfield  {journal} {\bibinfo  {journal} {Science}\ }\textbf {\bibinfo
  {volume} {305}},\ \bibinfo {pages} {1125} (\bibinfo {year} {2004})},\ \Eprint
  {https://arxiv.org/abs/https://www.science.org/doi/pdf/10.1126/science.1100700}
  {https://www.science.org/doi/pdf/10.1126/science.1100700} \BibitemShut
  {NoStop}%
\bibitem [{\citenamefont {Meinert}\ \emph {et~al.}(2015)\citenamefont
  {Meinert}, \citenamefont {Panfil}, \citenamefont {Mark}, \citenamefont
  {Lauber}, \citenamefont {Caux},\ and\ \citenamefont
  {N\"agerl}}]{PhysRevLett.115.085301}%
  \BibitemOpen
  \bibfield  {author} {\bibinfo {author} {\bibfnamefont {F.}~\bibnamefont
  {Meinert}}, \bibinfo {author} {\bibfnamefont {M.}~\bibnamefont {Panfil}},
  \bibinfo {author} {\bibfnamefont {M.~J.}\ \bibnamefont {Mark}}, \bibinfo
  {author} {\bibfnamefont {K.}~\bibnamefont {Lauber}}, \bibinfo {author}
  {\bibfnamefont {J.-S.}\ \bibnamefont {Caux}},\ and\ \bibinfo {author}
  {\bibfnamefont {H.-C.}\ \bibnamefont {N\"agerl}},\ }\bibfield  {title}
  {\bibinfo {title} {Probing the excitations of a lieb-liniger gas from weak to
  strong coupling},\ }\href {https://doi.org/10.1103/PhysRevLett.115.085301}
  {\bibfield  {journal} {\bibinfo  {journal} {Phys. Rev. Lett.}\ }\textbf
  {\bibinfo {volume} {115}},\ \bibinfo {pages} {085301} (\bibinfo {year}
  {2015})}\BibitemShut {NoStop}%
\bibitem [{\citenamefont {Moritz}\ \emph {et~al.}(2003)\citenamefont {Moritz},
  \citenamefont {St\"oferle}, \citenamefont {K\"ohl},\ and\ \citenamefont
  {Esslinger}}]{PhysRevLett.91.250402}%
  \BibitemOpen
  \bibfield  {author} {\bibinfo {author} {\bibfnamefont {H.}~\bibnamefont
  {Moritz}}, \bibinfo {author} {\bibfnamefont {T.}~\bibnamefont {St\"oferle}},
  \bibinfo {author} {\bibfnamefont {M.}~\bibnamefont {K\"ohl}},\ and\ \bibinfo
  {author} {\bibfnamefont {T.}~\bibnamefont {Esslinger}},\ }\bibfield  {title}
  {\bibinfo {title} {Exciting collective oscillations in a trapped 1d gas},\
  }\href {https://doi.org/10.1103/PhysRevLett.91.250402} {\bibfield  {journal}
  {\bibinfo  {journal} {Phys. Rev. Lett.}\ }\textbf {\bibinfo {volume} {91}},\
  \bibinfo {pages} {250402} (\bibinfo {year} {2003})}\BibitemShut {NoStop}%
\bibitem [{\citenamefont {De~Nardis}\ \emph {et~al.}(2018)\citenamefont
  {De~Nardis}, \citenamefont {Bernard},\ and\ \citenamefont
  {Doyon}}]{PhysRevLett.121.160603}%
  \BibitemOpen
  \bibfield  {author} {\bibinfo {author} {\bibfnamefont {J.}~\bibnamefont
  {De~Nardis}}, \bibinfo {author} {\bibfnamefont {D.}~\bibnamefont {Bernard}},\
  and\ \bibinfo {author} {\bibfnamefont {B.}~\bibnamefont {Doyon}},\ }\bibfield
   {title} {\bibinfo {title} {Hydrodynamic diffusion in integrable systems},\
  }\href {https://doi.org/10.1103/PhysRevLett.121.160603} {\bibfield  {journal}
  {\bibinfo  {journal} {Phys. Rev. Lett.}\ }\textbf {\bibinfo {volume} {121}},\
  \bibinfo {pages} {160603} (\bibinfo {year} {2018})}\BibitemShut {NoStop}%
\bibitem [{\citenamefont {Schemmer}\ \emph {et~al.}(2019)\citenamefont
  {Schemmer}, \citenamefont {Bouchoule}, \citenamefont {Doyon},\ and\
  \citenamefont {Dubail}}]{PhysRevLett.122.090601}%
  \BibitemOpen
  \bibfield  {author} {\bibinfo {author} {\bibfnamefont {M.}~\bibnamefont
  {Schemmer}}, \bibinfo {author} {\bibfnamefont {I.}~\bibnamefont {Bouchoule}},
  \bibinfo {author} {\bibfnamefont {B.}~\bibnamefont {Doyon}},\ and\ \bibinfo
  {author} {\bibfnamefont {J.}~\bibnamefont {Dubail}},\ }\bibfield  {title}
  {\bibinfo {title} {Generalized hydrodynamics on an atom chip},\ }\href
  {https://doi.org/10.1103/PhysRevLett.122.090601} {\bibfield  {journal}
  {\bibinfo  {journal} {Phys. Rev. Lett.}\ }\textbf {\bibinfo {volume} {122}},\
  \bibinfo {pages} {090601} (\bibinfo {year} {2019})}\BibitemShut {NoStop}%
\bibitem [{\citenamefont {Panfil}\ \emph {et~al.}(2022)\citenamefont {Panfil},
  \citenamefont {Gopalakrishnan},\ and\ \citenamefont {Konik}}]{quasi_1d}%
  \BibitemOpen
  \bibfield  {author} {\bibinfo {author} {\bibfnamefont {M.}~\bibnamefont
  {Panfil}}, \bibinfo {author} {\bibfnamefont {S.}~\bibnamefont
  {Gopalakrishnan}},\ and\ \bibinfo {author} {\bibfnamefont {R.~M.}\
  \bibnamefont {Konik}},\ }\href {https://arxiv.org/abs/2205.06492} {\bibinfo
  {title} {Thermalization of interacting quasi-one-dimensional systems}},\
  \bibinfo {howpublished} {(to appear in Phys. Rev. Lett., arXiv 2205.06492)}
  (\bibinfo {year} {2022})\BibitemShut {NoStop}%
\bibitem [{\citenamefont {Lenard}(1964)}]{Lenard64}%
  \BibitemOpen
  \bibfield  {author} {\bibinfo {author} {\bibfnamefont {A.}~\bibnamefont
  {Lenard}},\ }\bibfield  {title} {\bibinfo {title} {{Momentum Distribution in
  the Ground State of the One-Dimensional system of Impenetrable Bosons}},\
  }\href {https://doi.org/10.1063/1.1704196} {\bibfield  {journal} {\bibinfo
  {journal} {J. Math. Phys.}\ }\textbf {\bibinfo {volume} {5}},\ \bibinfo
  {pages} {930} (\bibinfo {year} {1964})}\BibitemShut {NoStop}%
\bibitem [{\citenamefont {Lenard}(1966)}]{Lenard66}%
  \BibitemOpen
  \bibfield  {author} {\bibinfo {author} {\bibfnamefont {A.}~\bibnamefont
  {Lenard}},\ }\bibfield  {title} {\bibinfo {title} {{One-Dimensional
  Impenetrable Bosons in Thermal Equilibrium}},\ }\href
  {https://doi.org/10.1063/1.1705029} {\bibfield  {journal} {\bibinfo
  {journal} {J. Math. Phys.}\ }\textbf {\bibinfo {volume} {7}},\ \bibinfo
  {pages} {1268} (\bibinfo {year} {1966})}\BibitemShut {NoStop}%
\bibitem [{\citenamefont {Nagamiya}(1940)}]{1940705}%
  \BibitemOpen
  \bibfield  {author} {\bibinfo {author} {\bibfnamefont {T.}~\bibnamefont
  {Nagamiya}},\ }\bibfield  {title} {\bibinfo {title} {Statistical mechanics of
  one-dimensional substances i},\ }\href
  {https://doi.org/10.11429/ppmsj1919.22.8-9_705} {\bibfield  {journal}
  {\bibinfo  {journal} {Proc. Phys.-Math. Soc. Japan}\ }\textbf {\bibinfo
  {volume} {22}},\ \bibinfo {pages} {705} (\bibinfo {year} {1940})}\BibitemShut
  {NoStop}%
\bibitem [{\citenamefont {Sutherland}(1971)}]{10.1063/1.1665585}%
  \BibitemOpen
  \bibfield  {author} {\bibinfo {author} {\bibfnamefont {B.}~\bibnamefont
  {Sutherland}},\ }\bibfield  {title} {\bibinfo {title} {Quantum many‐body
  problem in one dimension: Thermodynamics},\ }\href
  {https://doi.org/10.1063/1.1665585} {\bibfield  {journal} {\bibinfo
  {journal} {J. Math. Phys.}\ }\textbf {\bibinfo {volume} {12}},\ \bibinfo
  {pages} {251} (\bibinfo {year} {1971})}\BibitemShut {NoStop}%
\bibitem [{\citenamefont {Tonks}(1936)}]{PhysRev.50.955}%
  \BibitemOpen
  \bibfield  {author} {\bibinfo {author} {\bibfnamefont {L.}~\bibnamefont
  {Tonks}},\ }\bibfield  {title} {\bibinfo {title} {The complete equation of
  state of one, two and three-dimensional gases of hard elastic spheres},\
  }\href {https://doi.org/10.1103/PhysRev.50.955} {\bibfield  {journal}
  {\bibinfo  {journal} {Phys. Rev.}\ }\textbf {\bibinfo {volume} {50}},\
  \bibinfo {pages} {955} (\bibinfo {year} {1936})}\BibitemShut {NoStop}%
\bibitem [{\citenamefont {Salsburg}\ \emph {et~al.}(1953)\citenamefont
  {Salsburg}, \citenamefont {Zwanzig},\ and\ \citenamefont
  {Kirkwood}}]{SalsburgZwanzigKirkwood1953}%
  \BibitemOpen
  \bibfield  {author} {\bibinfo {author} {\bibfnamefont {Z.~W.}\ \bibnamefont
  {Salsburg}}, \bibinfo {author} {\bibfnamefont {R.~W.}\ \bibnamefont
  {Zwanzig}},\ and\ \bibinfo {author} {\bibfnamefont {J.~G.}\ \bibnamefont
  {Kirkwood}},\ }\bibfield  {title} {\bibinfo {title} {Molecular distribution
  functions in a one-dimensional fluid},\ }\href
  {https://doi.org/10.1063/1.1699116} {\bibfield  {journal} {\bibinfo
  {journal} {The Journal of Chemical Physics}\ }\textbf {\bibinfo {volume}
  {21}},\ \bibinfo {pages} {1098} (\bibinfo {year} {1953})}\BibitemShut
  {NoStop}%
\bibitem [{\citenamefont {Flicker}(1968)}]{10.1063/1.1664471}%
  \BibitemOpen
  \bibfield  {author} {\bibinfo {author} {\bibfnamefont {M.}~\bibnamefont
  {Flicker}},\ }\bibfield  {title} {\bibinfo {title} {Pair distribution
  function of a one‐dimensional hard rod gas},\ }\href
  {https://doi.org/10.1063/1.1664471} {\bibfield  {journal} {\bibinfo
  {journal} {Journal of Mathematical Physics}\ }\textbf {\bibinfo {volume}
  {9}},\ \bibinfo {pages} {171} (\bibinfo {year} {1968})}\BibitemShut {NoStop}%
\bibitem [{\citenamefont {Doyon}(2018)}]{Doyon2018_PRL_SolitonGases}%
  \BibitemOpen
  \bibfield  {author} {\bibinfo {author} {\bibfnamefont {B.}~\bibnamefont
  {Doyon}},\ }\bibfield  {title} {\bibinfo {title} {Soliton gases and
  generalized hydrodynamics},\ }\href
  {https://doi.org/10.1103/PhysRevLett.120.045301} {\bibfield  {journal}
  {\bibinfo  {journal} {Physical Review Letters}\ }\textbf {\bibinfo {volume}
  {120}},\ \bibinfo {pages} {045301} (\bibinfo {year} {2018})}\BibitemShut
  {NoStop}%
\bibitem [{\citenamefont {Doyon}\ and\ \citenamefont
  {Spohn}(2017)}]{DoyonSpohn2017_DomainWall}%
  \BibitemOpen
  \bibfield  {author} {\bibinfo {author} {\bibfnamefont {B.}~\bibnamefont
  {Doyon}}\ and\ \bibinfo {author} {\bibfnamefont {H.}~\bibnamefont {Spohn}},\
  }\bibfield  {title} {\bibinfo {title} {Dynamics of hard rods with initial
  domain wall state},\ }\href {https://doi.org/10.1088/1742-5468/aa7abf}
  {\bibfield  {journal} {\bibinfo  {journal} {Journal of Statistical Mechanics:
  Theory and Experiment}\ }\textbf {\bibinfo {volume} {2017}},\ \bibinfo
  {pages} {073210} (\bibinfo {year} {2017})}\BibitemShut {NoStop}%
\bibitem [{\citenamefont {Singh}\ \emph {et~al.}(2024)\citenamefont {Singh},
  \citenamefont {Dhar}, \citenamefont {Spohn},\ and\ \citenamefont
  {Kundu}}]{SinghDharSpohnKundu2024_JStatPhys_HardRods}%
  \BibitemOpen
  \bibfield  {author} {\bibinfo {author} {\bibfnamefont {S.~K.}\ \bibnamefont
  {Singh}}, \bibinfo {author} {\bibfnamefont {A.}~\bibnamefont {Dhar}},
  \bibinfo {author} {\bibfnamefont {H.}~\bibnamefont {Spohn}},\ and\ \bibinfo
  {author} {\bibfnamefont {A.}~\bibnamefont {Kundu}},\ }\bibfield  {title}
  {\bibinfo {title} {Thermalization and hydrodynamics in an interacting
  integrable system: The case of hard rods},\ }\href
  {https://doi.org/10.1007/s10955-024-03282-z} {\bibfield  {journal} {\bibinfo
  {journal} {Journal of Statistical Physics}\ }\textbf {\bibinfo {volume}
  {191}},\ \bibinfo {pages} {66} (\bibinfo {year} {2024})}\BibitemShut
  {NoStop}%
\bibitem [{\citenamefont {Spohn}(1991)}]{SpohnBOOK}%
  \BibitemOpen
  \bibfield  {author} {\bibinfo {author} {\bibfnamefont {H.}~\bibnamefont
  {Spohn}},\ }\bibfield  {title} {\bibinfo {title} {Large scale dynamics of
  interacting particles},\ }\href@noop {} {\bibfield  {journal} {\bibinfo
  {journal} {Springer, Berlin}\ } (\bibinfo {year} {1991})}\BibitemShut
  {NoStop}%
\bibitem [{\citenamefont {Motta}\ \emph {et~al.}(2016)\citenamefont {Motta},
  \citenamefont {Vitali}, \citenamefont {Rossi}, \citenamefont {Galli},\ and\
  \citenamefont {Bertaina}}]{PhysRevA.94.043627}%
  \BibitemOpen
  \bibfield  {author} {\bibinfo {author} {\bibfnamefont {M.}~\bibnamefont
  {Motta}}, \bibinfo {author} {\bibfnamefont {E.}~\bibnamefont {Vitali}},
  \bibinfo {author} {\bibfnamefont {M.}~\bibnamefont {Rossi}}, \bibinfo
  {author} {\bibfnamefont {D.~E.}\ \bibnamefont {Galli}},\ and\ \bibinfo
  {author} {\bibfnamefont {G.}~\bibnamefont {Bertaina}},\ }\bibfield  {title}
  {\bibinfo {title} {Dynamical structure factor of one-dimensional hard rods},\
  }\href {https://doi.org/10.1103/PhysRevA.94.043627} {\bibfield  {journal}
  {\bibinfo  {journal} {Phys. Rev. A}\ }\textbf {\bibinfo {volume} {94}},\
  \bibinfo {pages} {043627} (\bibinfo {year} {2016})}\BibitemShut {NoStop}%
\bibitem [{\citenamefont {Mazzanti}\ \emph
  {et~al.}(2008{\natexlab{a}})\citenamefont {Mazzanti}, \citenamefont
  {Astrakharchik}, \citenamefont {Boronat},\ and\ \citenamefont
  {Casulleras}}]{PhysRevA.77.043632}%
  \BibitemOpen
  \bibfield  {author} {\bibinfo {author} {\bibfnamefont {F.}~\bibnamefont
  {Mazzanti}}, \bibinfo {author} {\bibfnamefont {G.~E.}\ \bibnamefont
  {Astrakharchik}}, \bibinfo {author} {\bibfnamefont {J.}~\bibnamefont
  {Boronat}},\ and\ \bibinfo {author} {\bibfnamefont {J.}~\bibnamefont
  {Casulleras}},\ }\bibfield  {title} {\bibinfo {title} {Off-diagonal
  ground-state properties of a one-dimensional gas of fermi hard rods},\ }\href
  {https://doi.org/10.1103/PhysRevA.77.043632} {\bibfield  {journal} {\bibinfo
  {journal} {Phys. Rev. A}\ }\textbf {\bibinfo {volume} {77}},\ \bibinfo
  {pages} {043632} (\bibinfo {year} {2008}{\natexlab{a}})}\BibitemShut
  {NoStop}%
\bibitem [{\citenamefont {Mazzanti}\ \emph
  {et~al.}(2008{\natexlab{b}})\citenamefont {Mazzanti}, \citenamefont
  {Astrakharchik}, \citenamefont {Boronat},\ and\ \citenamefont
  {Casulleras}}]{PhysRevLett.100.020401}%
  \BibitemOpen
  \bibfield  {author} {\bibinfo {author} {\bibfnamefont {F.}~\bibnamefont
  {Mazzanti}}, \bibinfo {author} {\bibfnamefont {G.~E.}\ \bibnamefont
  {Astrakharchik}}, \bibinfo {author} {\bibfnamefont {J.}~\bibnamefont
  {Boronat}},\ and\ \bibinfo {author} {\bibfnamefont {J.}~\bibnamefont
  {Casulleras}},\ }\bibfield  {title} {\bibinfo {title} {Ground-state
  properties of a one-dimensional system of hard rods},\ }\href
  {https://doi.org/10.1103/PhysRevLett.100.020401} {\bibfield  {journal}
  {\bibinfo  {journal} {Phys. Rev. Lett.}\ }\textbf {\bibinfo {volume} {100}},\
  \bibinfo {pages} {020401} (\bibinfo {year} {2008}{\natexlab{b}})}\BibitemShut
  {NoStop}%
\bibitem [{\citenamefont {Pearce}\ \emph {et~al.}(2005)\citenamefont {Pearce},
  \citenamefont {Adams}, \citenamefont {Vilches}, \citenamefont {Johnson},\
  and\ \citenamefont {Glyde}}]{PhysRevLett.95.185302}%
  \BibitemOpen
  \bibfield  {author} {\bibinfo {author} {\bibfnamefont {J.~V.}\ \bibnamefont
  {Pearce}}, \bibinfo {author} {\bibfnamefont {M.~A.}\ \bibnamefont {Adams}},
  \bibinfo {author} {\bibfnamefont {O.~E.}\ \bibnamefont {Vilches}}, \bibinfo
  {author} {\bibfnamefont {M.~R.}\ \bibnamefont {Johnson}},\ and\ \bibinfo
  {author} {\bibfnamefont {H.~R.}\ \bibnamefont {Glyde}},\ }\bibfield  {title}
  {\bibinfo {title} {One-dimensional and two-dimensional quantum systems on
  carbon nanotube bundles},\ }\href
  {https://doi.org/10.1103/PhysRevLett.95.185302} {\bibfield  {journal}
  {\bibinfo  {journal} {Phys. Rev. Lett.}\ }\textbf {\bibinfo {volume} {95}},\
  \bibinfo {pages} {185302} (\bibinfo {year} {2005})}\BibitemShut {NoStop}%
\bibitem [{\citenamefont {Yu}\ \emph {et~al.}(2025)\citenamefont {Yu},
  \citenamefont {Zhu},\ and\ \citenamefont
  {Sanchez-Palencia}}]{yu2025exactsolutionluttingerliquid}%
  \BibitemOpen
  \bibfield  {author} {\bibinfo {author} {\bibfnamefont {S.}~\bibnamefont
  {Yu}}, \bibinfo {author} {\bibfnamefont {Z.}~\bibnamefont {Zhu}},\ and\
  \bibinfo {author} {\bibfnamefont {L.}~\bibnamefont {Sanchez-Palencia}},\
  }\href {https://arxiv.org/abs/2505.20376} {\bibinfo {title} {Exact solution
  and luttinger liquid behavior of the quantum 1d hard rod model}} (\bibinfo
  {year} {2025}),\ \Eprint {https://arxiv.org/abs/2505.20376} {arXiv:2505.20376
  [cond-mat.str-el]} \BibitemShut {NoStop}%
\bibitem [{\citenamefont {Bethe}(1931)}]{Bethe1931}%
  \BibitemOpen
  \bibfield  {author} {\bibinfo {author} {\bibfnamefont {H.}~\bibnamefont
  {Bethe}},\ }\bibfield  {title} {\bibinfo {title} {{Zur Theorie der Metalle:
  I. Eigenwerte und Eigenfunktionen der linearen Atomkette}},\ }\href
  {https://doi.org/10.1007/BF01341708} {\bibfield  {journal} {\bibinfo
  {journal} {Zeitschrift f{\"u}r Physik}\ }\textbf {\bibinfo {volume} {71}},\
  \bibinfo {pages} {205} (\bibinfo {year} {1931})}\BibitemShut {NoStop}%
\bibitem [{\citenamefont {Giamarchi}(2004{\natexlab{b}})}]{GiamarchiBOOK}%
  \BibitemOpen
  \bibfield  {author} {\bibinfo {author} {\bibfnamefont {T.}~\bibnamefont
  {Giamarchi}},\ }\href@noop {} {\emph {\bibinfo {title} {Quantum Physics in
  One Dimension}}}\ (\bibinfo  {publisher} {Oxford University Press},\ \bibinfo
  {year} {2004})\BibitemShut {NoStop}%
\bibitem [{\citenamefont {Lieb}\ and\ \citenamefont
  {Liniger}(1963)}]{1963_Lieb_PR_130_1}%
  \BibitemOpen
  \bibfield  {author} {\bibinfo {author} {\bibfnamefont {E.~H.}\ \bibnamefont
  {Lieb}}\ and\ \bibinfo {author} {\bibfnamefont {W.}~\bibnamefont {Liniger}},\
  }\bibfield  {title} {\bibinfo {title} {{Exact Analysis of an Interacting Bose
  Gas. I. The General Solution and the Ground State}},\ }\href
  {https://doi.org/10.1103/PhysRev.130.1605} {\bibfield  {journal} {\bibinfo
  {journal} {Phys. Rev.}\ }\textbf {\bibinfo {volume} {130}},\ \bibinfo {pages}
  {1605} (\bibinfo {year} {1963})}\BibitemShut {NoStop}%
\bibitem [{\citenamefont {Slavnov}(1990)}]{1990_Slavnov_TMP_82}%
  \BibitemOpen
  \bibfield  {author} {\bibinfo {author} {\bibfnamefont {N.~A.}\ \bibnamefont
  {Slavnov}},\ }\bibfield  {title} {\bibinfo {title} {Nonequal-time current
  correlation function in a one-dimensional bose gas},\ }\href@noop {}
  {\bibfield  {journal} {\bibinfo  {journal} {Theor. Math. Phys.}\ }\textbf
  {\bibinfo {volume} {82}},\ \bibinfo {pages} {273} (\bibinfo {year}
  {1990})}\BibitemShut {NoStop}%
\bibitem [{\citenamefont {Essler}\ and\ \citenamefont
  {de~Klerk}(2024)}]{PhysRevX.14.031048}%
  \BibitemOpen
  \bibfield  {author} {\bibinfo {author} {\bibfnamefont {F.~H.~L.}\
  \bibnamefont {Essler}}\ and\ \bibinfo {author} {\bibfnamefont {A.~J. J.~M.}\
  \bibnamefont {de~Klerk}},\ }\bibfield  {title} {\bibinfo {title} {Statistics
  of matrix elements of local operators in integrable models},\ }\href
  {https://doi.org/10.1103/PhysRevX.14.031048} {\bibfield  {journal} {\bibinfo
  {journal} {Phys. Rev. X}\ }\textbf {\bibinfo {volume} {14}},\ \bibinfo
  {pages} {031048} (\bibinfo {year} {2024})}\BibitemShut {NoStop}%
\bibitem [{\citenamefont {Caux}\ and\ \citenamefont
  {Maillet}(2005)}]{PhysRevLett.95.077201}%
  \BibitemOpen
  \bibfield  {author} {\bibinfo {author} {\bibfnamefont {J.-S.}\ \bibnamefont
  {Caux}}\ and\ \bibinfo {author} {\bibfnamefont {J.~M.}\ \bibnamefont
  {Maillet}},\ }\bibfield  {title} {\bibinfo {title} {Computation of dynamical
  correlation functions of heisenberg chains in a magnetic field},\ }\href
  {https://doi.org/10.1103/PhysRevLett.95.077201} {\bibfield  {journal}
  {\bibinfo  {journal} {Phys. Rev. Lett.}\ }\textbf {\bibinfo {volume} {95}},\
  \bibinfo {pages} {077201} (\bibinfo {year} {2005})}\BibitemShut {NoStop}%
\bibitem [{\citenamefont {Caux}\ and\ \citenamefont
  {Calabrese}(2006)}]{2006_Caux_PRA_74}%
  \BibitemOpen
  \bibfield  {author} {\bibinfo {author} {\bibfnamefont {J.-S.}\ \bibnamefont
  {Caux}}\ and\ \bibinfo {author} {\bibfnamefont {P.}~\bibnamefont
  {Calabrese}},\ }\bibfield  {title} {\bibinfo {title} {Dynamical
  density-density correlations in the one-dimensional {B}ose gas},\ }\href
  {https://doi.org/10.1103/PhysRevA.74.031605} {\bibfield  {journal} {\bibinfo
  {journal} {Phys. Rev. A}\ }\textbf {\bibinfo {volume} {74}},\ \bibinfo
  {pages} {031605} (\bibinfo {year} {2006})}\BibitemShut {NoStop}%
\bibitem [{\citenamefont {Caux}\ \emph {et~al.}(2007)\citenamefont {Caux},
  \citenamefont {Calabrese},\ and\ \citenamefont
  {Slavnov}}]{1742-5468-2007-01-P01008}%
  \BibitemOpen
  \bibfield  {author} {\bibinfo {author} {\bibfnamefont {J.-S.}\ \bibnamefont
  {Caux}}, \bibinfo {author} {\bibfnamefont {P.}~\bibnamefont {Calabrese}},\
  and\ \bibinfo {author} {\bibfnamefont {N.~A.}\ \bibnamefont {Slavnov}},\
  }\bibfield  {title} {\bibinfo {title} {One-particle dynamical correlations in
  the one-dimensional bose gas},\ }\href
  {https://doi.org/10.1088/1742-5468/2007/01/P01008} {\bibfield  {journal}
  {\bibinfo  {journal} {J. Stat. Mech. Theor. Exp.}\ }\textbf {\bibinfo
  {volume} {2007}},\ \bibinfo {pages} {P01008} (\bibinfo {year}
  {2007})}\BibitemShut {NoStop}%
\bibitem [{\citenamefont {{Caux}}(2009)}]{2009_Caux_JMP_50}%
  \BibitemOpen
  \bibfield  {author} {\bibinfo {author} {\bibfnamefont {J.-S.}\ \bibnamefont
  {{Caux}}},\ }\bibfield  {title} {\bibinfo {title} {{Correlation functions of
  integrable models: A description of the ABACUS algorithm}},\ }\href
  {https://doi.org/10.1063/1.3216474} {\bibfield  {journal} {\bibinfo
  {journal} {J. Math. Phys.}\ }\textbf {\bibinfo {volume} {50}},\ \bibinfo
  {pages} {095214} (\bibinfo {year} {2009})}\BibitemShut {NoStop}%
\bibitem [{\citenamefont {Li}\ \emph {et~al.}(2023)\citenamefont {Li},
  \citenamefont {Cheng}, \citenamefont {Chen},\ and\ \citenamefont
  {Guan}}]{Li_2023}%
  \BibitemOpen
  \bibfield  {author} {\bibinfo {author} {\bibfnamefont {R.-T.}\ \bibnamefont
  {Li}}, \bibinfo {author} {\bibfnamefont {S.}~\bibnamefont {Cheng}}, \bibinfo
  {author} {\bibfnamefont {Y.-Y.}\ \bibnamefont {Chen}},\ and\ \bibinfo
  {author} {\bibfnamefont {X.-W.}\ \bibnamefont {Guan}},\ }\bibfield  {title}
  {\bibinfo {title} {Exact results of dynamical structure factor of
  lieb--liniger model},\ }\href {https://doi.org/10.1088/1751-8121/ace80f}
  {\bibfield  {journal} {\bibinfo  {journal} {Journal of Physics A:
  Mathematical and Theoretical}\ }\textbf {\bibinfo {volume} {56}},\ \bibinfo
  {pages} {335204} (\bibinfo {year} {2023})}\BibitemShut {NoStop}%
\bibitem [{\citenamefont {Caux}()}]{ABACUS}%
  \BibitemOpen
  \bibfield  {author} {\bibinfo {author} {\bibfnamefont {J.-S.}\ \bibnamefont
  {Caux}},\ }\href@noop {} {\bibinfo {title} {{ABACUS} (version v1), general
  set of algorithms for dealing with {Bethe Ansatz-solvable} systems}},\
  \Eprint {https://arxiv.org/abs/https://git.jscaux.org/jscaux/ABACUS-v1}
  {https://git.jscaux.org/jscaux/ABACUS-v1} \BibitemShut {NoStop}%
\bibitem [{\citenamefont {Imambekov}\ and\ \citenamefont
  {Glazman}(2008)}]{Imambekov_2008}%
  \BibitemOpen
  \bibfield  {author} {\bibinfo {author} {\bibfnamefont {A.}~\bibnamefont
  {Imambekov}}\ and\ \bibinfo {author} {\bibfnamefont {L.~I.}\ \bibnamefont
  {Glazman}},\ }\bibfield  {title} {\bibinfo {title} {Exact exponents of edge
  singularities in dynamic correlation functions of 1d bose gas},\ }\bibfield
  {journal} {\bibinfo  {journal} {Physical Review Letters}\ }\textbf {\bibinfo
  {volume} {100}},\ \href {https://doi.org/10.1103/physrevlett.100.206805}
  {10.1103/physrevlett.100.206805} (\bibinfo {year} {2008})\BibitemShut
  {NoStop}%
\bibitem [{\citenamefont {Imambekov}\ and\ \citenamefont
  {Glazman}(2009)}]{Imambekov_2009}%
  \BibitemOpen
  \bibfield  {author} {\bibinfo {author} {\bibfnamefont {A.}~\bibnamefont
  {Imambekov}}\ and\ \bibinfo {author} {\bibfnamefont {L.~I.}\ \bibnamefont
  {Glazman}},\ }\bibfield  {title} {\bibinfo {title} {Universal theory of
  nonlinear luttinger liquids},\ }\href
  {https://doi.org/10.1126/science.1165403} {\bibfield  {journal} {\bibinfo
  {journal} {Science}\ }\textbf {\bibinfo {volume} {323}},\ \bibinfo {pages}
  {228} (\bibinfo {year} {2009})}\BibitemShut {NoStop}%
\bibitem [{\citenamefont {Panfil}\ \emph {et~al.}(2013)\citenamefont {Panfil},
  \citenamefont {De~Nardis},\ and\ \citenamefont
  {Caux}}]{PhysRevLett.110.125302}%
  \BibitemOpen
  \bibfield  {author} {\bibinfo {author} {\bibfnamefont {M.}~\bibnamefont
  {Panfil}}, \bibinfo {author} {\bibfnamefont {J.}~\bibnamefont {De~Nardis}},\
  and\ \bibinfo {author} {\bibfnamefont {J.-S.}\ \bibnamefont {Caux}},\
  }\bibfield  {title} {\bibinfo {title} {Metastable criticality and the super
  tonks-girardeau gas},\ }\href
  {https://doi.org/10.1103/PhysRevLett.110.125302} {\bibfield  {journal}
  {\bibinfo  {journal} {Phys. Rev. Lett.}\ }\textbf {\bibinfo {volume} {110}},\
  \bibinfo {pages} {125302} (\bibinfo {year} {2013})}\BibitemShut {NoStop}%
\bibitem [{\citenamefont {Shashi}\ \emph {et~al.}(2011)\citenamefont {Shashi},
  \citenamefont {Glazman}, \citenamefont {Caux},\ and\ \citenamefont
  {Imambekov}}]{2011_Shashi_PRB_84}%
  \BibitemOpen
  \bibfield  {author} {\bibinfo {author} {\bibfnamefont {A.}~\bibnamefont
  {Shashi}}, \bibinfo {author} {\bibfnamefont {L.~I.}\ \bibnamefont {Glazman}},
  \bibinfo {author} {\bibfnamefont {J.-S.}\ \bibnamefont {Caux}},\ and\
  \bibinfo {author} {\bibfnamefont {A.}~\bibnamefont {Imambekov}},\ }\bibfield
  {title} {\bibinfo {title} {Nonuniversal prefactors in the correlation
  functions of one-dimensional quantum liquids},\ }\href
  {https://doi.org/10.1103/PhysRevB.84.045408} {\bibfield  {journal} {\bibinfo
  {journal} {Phys. Rev. B}\ }\textbf {\bibinfo {volume} {84}},\ \bibinfo
  {pages} {045408} (\bibinfo {year} {2011})}\BibitemShut {NoStop}%
\bibitem [{\citenamefont {Shashi}\ \emph {et~al.}(2012)\citenamefont {Shashi},
  \citenamefont {Panfil}, \citenamefont {Caux},\ and\ \citenamefont
  {Imambekov}}]{Shashi_2012}%
  \BibitemOpen
  \bibfield  {author} {\bibinfo {author} {\bibfnamefont {A.}~\bibnamefont
  {Shashi}}, \bibinfo {author} {\bibfnamefont {M.}~\bibnamefont {Panfil}},
  \bibinfo {author} {\bibfnamefont {J.-S.}\ \bibnamefont {Caux}},\ and\
  \bibinfo {author} {\bibfnamefont {A.}~\bibnamefont {Imambekov}},\ }\bibfield
  {title} {\bibinfo {title} {Exact prefactors in static and dynamic correlation
  functions of one-dimensional quantum integrable models: Applications to the
  {Calogero-Sutherland}, {Lieb-Liniger}, and {XXZ} models},\ }\bibfield
  {journal} {\bibinfo  {journal} {Physical Review B}\ }\textbf {\bibinfo
  {volume} {85}},\ \href {https://doi.org/10.1103/physrevb.85.155136}
  {10.1103/physrevb.85.155136} (\bibinfo {year} {2012})\BibitemShut {NoStop}%
\bibitem [{\citenamefont {Pozsgay}\ \emph
  {et~al.}(2021{\natexlab{a}})\citenamefont {Pozsgay}, \citenamefont {Gombor},
  \citenamefont {Hutsalyuk}, \citenamefont {Jiang}, \citenamefont
  {Pristy\'ak},\ and\ \citenamefont {Vernier}}]{PhysRevE.104.044106}%
  \BibitemOpen
  \bibfield  {author} {\bibinfo {author} {\bibfnamefont {B.}~\bibnamefont
  {Pozsgay}}, \bibinfo {author} {\bibfnamefont {T.}~\bibnamefont {Gombor}},
  \bibinfo {author} {\bibfnamefont {A.}~\bibnamefont {Hutsalyuk}}, \bibinfo
  {author} {\bibfnamefont {Y.}~\bibnamefont {Jiang}}, \bibinfo {author}
  {\bibfnamefont {L.}~\bibnamefont {Pristy\'ak}},\ and\ \bibinfo {author}
  {\bibfnamefont {E.}~\bibnamefont {Vernier}},\ }\bibfield  {title} {\bibinfo
  {title} {Integrable spin chain with hilbert space fragmentation and solvable
  real-time dynamics},\ }\href {https://doi.org/10.1103/PhysRevE.104.044106}
  {\bibfield  {journal} {\bibinfo  {journal} {Phys. Rev. E}\ }\textbf {\bibinfo
  {volume} {104}},\ \bibinfo {pages} {044106} (\bibinfo {year}
  {2021}{\natexlab{a}})}\BibitemShut {NoStop}%
\bibitem [{\citenamefont {Pozsgay}\ \emph
  {et~al.}(2021{\natexlab{b}})\citenamefont {Pozsgay}, \citenamefont {Gombor},\
  and\ \citenamefont {Hutsalyuk}}]{PhysRevE.104.064124}%
  \BibitemOpen
  \bibfield  {author} {\bibinfo {author} {\bibfnamefont {B.}~\bibnamefont
  {Pozsgay}}, \bibinfo {author} {\bibfnamefont {T.}~\bibnamefont {Gombor}},\
  and\ \bibinfo {author} {\bibfnamefont {A.}~\bibnamefont {Hutsalyuk}},\
  }\bibfield  {title} {\bibinfo {title} {Integrable hard-rod deformation of the
  heisenberg spin chains},\ }\href
  {https://doi.org/10.1103/PhysRevE.104.064124} {\bibfield  {journal} {\bibinfo
   {journal} {Phys. Rev. E}\ }\textbf {\bibinfo {volume} {104}},\ \bibinfo
  {pages} {064124} (\bibinfo {year} {2021}{\natexlab{b}})}\BibitemShut
  {NoStop}%
\bibitem [{\citenamefont {Cardy}\ and\ \citenamefont
  {Doyon}(2022)}]{Cardy:2022aa}%
  \BibitemOpen
  \bibfield  {author} {\bibinfo {author} {\bibfnamefont {J.}~\bibnamefont
  {Cardy}}\ and\ \bibinfo {author} {\bibfnamefont {B.}~\bibnamefont {Doyon}},\
  }\bibfield  {title} {\bibinfo {title} {{$\bar{T}T$} deformations and the
  width of fundamental particles},\ }\href
  {https://doi.org/10.1007/JHEP04(2022)136} {\bibfield  {journal} {\bibinfo
  {journal} {Journal of High Energy Physics}\ }\textbf {\bibinfo {volume}
  {2022}},\ \bibinfo {pages} {136} (\bibinfo {year} {2022})}\BibitemShut
  {NoStop}%
\bibitem [{\citenamefont {Panfil}\ \emph {et~al.}(2025)\citenamefont {Panfil},
  \citenamefont {Kiedrzy{\'n}ski},\ and\ \citenamefont {Witkowska}}]{dataQHR}%
  \BibitemOpen
  \bibfield  {author} {\bibinfo {author} {\bibfnamefont {M.}~\bibnamefont
  {Panfil}}, \bibinfo {author} {\bibfnamefont {S.}~\bibnamefont
  {Kiedrzy{\'n}ski}},\ and\ \bibinfo {author} {\bibfnamefont {E.}~\bibnamefont
  {Witkowska}},\ }\href {https://doi.org/10.58132/VQHEPL} {\bibinfo {title}
  {{Dynamic structure factor of quantum hard rods from exact form-factors}}}
  (\bibinfo {year} {2025})\BibitemShut {NoStop}%
\end{thebibliography}%

\appendix

\section{Slater determinant form of the wave-function} \label{app:Slater}

Let $\{r\}$ be an arbitrary set of $N$ ordered numbers: $\{r_1 < r_2 < \dots < r_N\}$ and let $\sigma \in \mathcal{P}_N$ be an arbitrary permutation of $N$ elements. Then
\begin{equation}
    \sum_{j=1}^N {\rm sgn}(r_{\sigma_j} - r_{\sigma_l}) = 2 \left(\sigma_j - \frac{N+1}{2} \right),
\end{equation}
where we adopt a convention that ${\rm sgn}(0) = 0$. 
This identity is useful to show a simplification to the phase term in the generic Bethe Ansatz wavefunction~\eqref{wave_function_perm} in the case of a linear phase shift. We have
\begin{equation}
    \frac{1}{2}\sum_{j>l}^N {\rm sgn}(r_j - r_l)\theta(\lambda_{\sigma_j} - \lambda_{\sigma_l}) = a \sum_j \lambda_{\sigma_j} \left(j - \frac{N+1}{2} \right).
\end{equation}
The wave function becomes then
\begin{equation}
    \psi_{\sigma}(\bfr| \bfl) = e^{i a P (N-1)/2} \times e^{i \sum_{j=1}^N \lambda_{\sigma_j} (r_j - a (j-1))}.
\end{equation}
The permutation independent prefactor sets a global phase of the wavefunction. This changes the form-factor~\eqref{form_factor} to
\begin{equation}
	\langle \bfm | \hat{\rho}(0) | \bfl \rangle = \frac{1}{L}\frac{1}{L_{\rm f}^{N-1}} \Delta P \left(2\sin \frac{a \Delta P}{2}\right)^{N-1} \mathcal{C}[\bfm, \bfl].
\end{equation}
Since the change involves just a phase factor it does not influence the correlation functions. For this reason we disregarded the phase factor in the main text.

\section{Computation of the TG gas form-factors} \label{app:TG_formfactors}
Analogously to Eq.~\eqref{eq: Formfactor}, the form factor of the TG gas can be written as  
\begin{equation}
	\langle \bfm | \hat{\rho}(0)| \bfl\rangle_{TG} = \frac{N}{N! L^N} \sum_{\sigma, \tau \in \mathcal{P}_N} (-1)^{|\sigma| + |\tau|} \mathcal{I}_{TG}(\sigma, \tau),
\end{equation}
where $\mathcal{I}_{TG}(\sigma, \tau)$ is defined by  
\begin{equation}
	\mathcal{I}_{TG}(\sigma, \tau) = \int {\rm d}{\bfr}_N \, \delta(r_1) \prod_{j=1}^N e^{i\left(\mu_{{\sigma}_j} - \lambda_{{\tau}_j} \right)r_j},
\end{equation}
with $\mu$ and $\lambda$ denoting the rapidities of the TG gas, given by $\mu_j \{\lambda_j\} = \frac{2\pi}{L} I_j$, where $I_j$ are integers.  

Using the quantization condition, the exponential factor in the expression above simplifies to  
\begin{equation} \label{app:form_factor_TG}
   e^{i\left(\mu_{{\sigma}_j} - \lambda_{{\tau}_j} \right)r_j} = e^{2i\pi\left(I_{{\sigma}_j} - I_{{\tau}_j} \right)r_j / L}.
\end{equation}

It follows that $\mathcal{I}_{TG}(\sigma, \tau)$ is nonzero if and only if the sets of quantum numbers (or equivalently rapidities $\bfm$ and $\bfl$) differ by at most a single element. In particular:
\begin{itemize}
    \item If $\bfm = \bfl$, there are $N!$ distinct pairs $(\sigma, \tau)$ yielding nonzero $\mathcal{I}_{TG}(\sigma, \tau)$.
    \item If $\bfm$ and $\bfl$ share exactly $N - 1$ elements, there are $(N-1)!$ such pairs.
\end{itemize}
From here, it is straightforward to see that \eqref{form_factor_TG} is indeed correct. 

The formula mentioned above can also be obtained by taking the limit $a \to 0$ of the hard-rod form factors. Consider the expression
\begin{equation}
   \eta_{\bfl,\bfm}= \Delta P\,\frac{(1-e^{i a \Delta P})^{N-1}}{\prod_{i,j}(\lambda_i-\mu_j)}.
\end{equation}
As $a \to 0$, the numerator vanishes asymptotically as $a^{\,N-1}$, while the denominator vanishes as $a^{\,n_{\bfl,\bfm}}$, where $n_{\bfl,\bfm}$ denotes the number of rapidities shared by the two states. Consequently, the Tonks--Girardeau form factors vanish whenever $n_{\bfl,\bfm} < N-1$.  

In the case $n_{\bfl,\bfm} = N-1$, the denominator reduces to
\begin{equation}
\prod_{i<j}(\lambda_j-\mu_i)(\lambda_j-\mu_i)\times (\lambda_- - \mu_+),
\end{equation}
with $\Delta P = \lambda_- - \mu_+$, which further simplifies the expression.  

If instead $\bfl=\bfm$, then $\Delta P=0$, and the numerator acquires an additional asymptotic suppression of order~$a$.  

Collecting these results, we obtain
\begin{align}
    \lim_{a\to0}\langle\bfm|\hat{\rho}(0)|\bfl\rangle
    &= \frac{1}{L^N}\,\lim_{a\to 0}\,
    \prod_{i<j} (\lambda_j-\lambda_i)(\mu_j-\mu_i)\,\eta_{\bfl,\bfm} \nonumber \\
    &= \frac{1}{L}\begin{cases}
       0, & n_{\bfl,\bfm} < N-1, \\[6pt]
       1, & n_{\bfl,\bfm} = N-1, \\[6pt]
       N, & n_{\bfl,\bfm} = N,
    \end{cases}
\end{align}
again in agreement with~\eqref{form_factor_TG}.

\section{Computation of the Luttinger liquid prefactors} \label{app:prefactor}

The rapidities of the $m$-th umklapp state are
\begin{equation}
	\mu_j = \lambda_j + \frac{2\pi}{L_{\rm f}} \nu_m,
\end{equation}
where $\nu_m = m \sqrt{K}$ and we used the Luttinger liquid parameter $K$. We write the form-factor as
\begin{equation}
    \langle \bfm | \hat{\rho}(0) | \bfl \rangle = M \times \mathcal{C}[\bfm, \bfl].
\end{equation}
where the prefactor (up to an irrelevant phase factor) is
\begin{equation}
	M = \frac{1}{L}\frac{1}{L_{\rm f}^{N-1}} \Delta P \left(2\sin \frac{a \Delta P}{2}\right)^{N-1}.
\end{equation}
For the umklapp excited state 
\begin{equation} \label{app:prefactor_umklapp}
    M = \frac{2k_F m }{L} \left(\frac{2\sin a k_F m}{L_{\rm f}}\right)^{N-1}.
\end{equation}

The determinant of the Cauchy matrix for the umklapp excited state is
\begin{equation}
	\det\left(\frac{1}{\lambda_i - \mu_j} \right) = \left(-\frac{L_f}{2\pi \nu_m}\right)^N \prod_{i \neq j}^N \frac{\lambda_i - \lambda_j}{\lambda_i - \lambda_j - \frac{2\pi}{L_{\rm f}}\nu_m}.
\end{equation}
The products can be expressed in terms of the quantum numbers. This gives
\begin{equation}
	\prod_{i \neq j}^N \frac{\lambda_i - \lambda_j}{\lambda_i - \lambda_j - \frac{2\pi}{L_{\rm f}}\nu_m} = \prod_{i \neq j}^N \frac{i - j}{i - j - \nu_m}.
\end{equation}
Furthermore, the products can be expressed with the help of the Gamma and Barnes-$G$ functions. We use the identity
\begin{equation}
	\prod_{i > j}^N (i-j - \nu) = \frac{1}{\Gamma^N(1-\nu)}\frac{G(N+1 -\nu)}{G(1-\nu)}.
\end{equation}
This gives
\begin{align}
	\prod_{i\neq j}^N (i-j) &= (-1)^{N(N-1)/2}G^2(N+1), \\
	\prod_{i\neq j}^N (i-j - \nu_m) &= (-1)^{N(N-1)/2}\left(\frac{\sin(\pi\nu_m)}{\pi \nu_m}\right)^N \nonumber \\
    &\times\frac{G(N+1+\nu_m)G(N+1-\nu_m)}{G(1+\nu_m)G(1-\nu_m)},
\end{align}
where in the second expression we used the Euler reflection formula
\begin{equation}
	\Gamma(1+z)\Gamma(1-z) = \frac{\pi z}{\sin(\pi z)}.
\end{equation}
The Cauchy determinant becomes
\begin{align} \label{app:cauchy_umklapp}
	\det\left(\frac{1}{\lambda_i - \mu_j} \right) &=  \left(\frac{L_f }{2 \sin(\pi \nu_m)}\right)^N \nonumber \\
	\times &\frac{G(1+\nu_m)G(1-\nu_m) G^2(N+1)}{G(N+1+\nu_m)G(N+1-\nu_m)}.
\end{align}
The sine term can be simplified by noting that $\nu_m = m \sqrt{K} = m(1 - a \rho_0)$ and using periodicity of the sine function. 
Combining this with the expression~\eqref{app:prefactor_umklapp} for the prefactor $M$, the absolute value of the form-factor reads
\begin{align}
    |\langle m, N | \hat{\rho}(0) | N \rangle| &= \frac{k_F m \sqrt{K} }{|\sin(a k_F m)|} \nonumber \\
    \times &\frac{G(1+\nu_m)G(1-\nu_m) G^2(N+1)}{G(N+1+\nu_m)G(N+1-\nu_m)},
\end{align}
as stated in the main text.

\end{document}